\documentclass[aps,prb,a4paper,superscriptaddress,preprint,showpacs]{revtex4}
\usepackage{graphicx}
\usepackage{graphicx,epsfig}
\usepackage{amsmath}
\input{comment.sty}
\begin{document}
\newcommand{\be}{\begin{equation}}
\newcommand{\ee}{\end{equation}}
\newcommand{\bea}{\begin{eqnarray}}
\newcommand{\eea}{\end{eqnarray}}
\newcommand{\ba}{\begin{eqnarray*}}
\newcommand{\ea}{\end{eqnarray*}}
\newcommand{\dagga}{{\phantom{\dagger}}}
\newcommand{\vect}[1]{\mathbf{#1}}
\newcommand{\A}{{\cal{A}}}
\title{Spectral properties of a two-orbital  
Anderson impurity model across a non-Fermi liquid fixed point}
\author{Lorenzo De Leo}  
\affiliation{International School for Advanced Studies (SISSA), and 
Istituto Nazionale per la Fisica della Materia (INFM) UR-Trieste SISSA, Via Beirut 2-4, 
I-34014 Trieste, Italy} 
\author{Michele Fabrizio} 
\affiliation{International School for Advanced Studies (SISSA), and Istituto 
Nazionale per la Fisica della Materia (INFM) UR-Trieste SISSA, Via Beirut 2-4, 
I-34014 Trieste, Italy} 
\affiliation{The Abdus Salam International Center for Theoretical Physics 
(ICTP), 
P.O.Box 586, I-34014 Trieste, Italy} 
\date{February 4, 2004} 
\begin{abstract} 
We study by Wilson numerical renormalization group the spectral properties 
of a two-orbital Anderson impurity model in the presence of an exchange splitting 
which follows either regular or inverted Hund's rules. 
The phase diagram contains a non-Fermi liquid fixed point separating a 
screened phase, where conventional Kondo effect occurs, from an 
unscreened one, where the exchange-splitting takes care of 
quenching the impurity degrees of freedom. 
On the Kondo screened side close to this fixed point the impurity density of states 
shows a narrow Kondo-peak on top of a broader resonance. This narrow 
peak transforms  in the unscreened phase 
into a narrow pseudo-gap inside the broad resonance.  
Right at the fixed point only the latter 
survives. The fixed point is therefore identified by a jump of the density of states 
at the chemical potential.  
We also consider the effect of several particle-hole symmetry-breaking terms. 
We show that particle-hole perturbations which simply shift the orbital energies 
do not wash out the fixed point, unlike those perturbations 
which hybridize the two orbitals. 
Consequently the density-of-state jump 
at the chemical potential remains finite even away from particle-hole symmetry.
In other words, the pseudo-gap stays pinned at the chemical potential, although it 
is partially filled in. 
We also discuss the relevance of these results for lattice models which map 
onto this Anderson impurity model in the limit of large lattice-coordination.    
Upon approaching the Mott metal-insulator transition, 
these lattice models necessarily enter a region with a   
local criticality which reflects the impurity non-Fermi liquid fixed point.    
However, unlike the impurity, the lattice can get rid of the 
single-impurity fixed-point instability by spontaneously developing bulk-coherent 
symmetry-broken phases, which we identify for different lattice models.

\end{abstract} 
 
\maketitle 

\section{Introduction}

Non-Fermi liquid behavior may emerge in Anderson and Kondo impurity 
models for two distinct reasons. The first one is that, by construction, the 
conduction electrons may not be able to perfectly Kondo-screen the impurity degrees of freedom. 
This is realized for instance in multi-channel Kondo models\cite{Nozieres&Blandin}. 

The alternative route towards non-Fermi liquid behavior is the presence of an 
intra-impurity mechanism which splits the impurity degeneracy favoring 
a non-degenerate configuration. 
The Kondo exchange takes advantage of letting the impurity tunnel among all available 
electronic configurations. This quantum tunneling is hampered by any term which splits the  
degeneracy and tends to trap the impurity into a given state. 
Therefore either the Kondo exchange overwhelms 
the intra-impurity splitting mechanism, or vice-versa, 
which leads respectively to a Kondo-screened phase  
or an unscreened one. When none of the two effects prevails, 
a non trivial behavior may appear. 
This is actually what happens in the two S=1/2-impurity Kondo model in the presence 
of an antiferromagnetic 
direct exchange between the impurity-spins\cite{J&V,A&L}. 
There it is known that, under particular circumstances\cite{AL&J}, 
an unstable non-Fermi liquid fixed point separates the Kondo-screened 
and unscreened regimes. 
Since this fixed point requires fine tuning of the model parameters, 
it is tempting to conclude that it is of little physical relevance. 
In reality a similar competition may be at the heart of 
strongly-correlated electron lattice-models. 
Here the kinetic energy profits by the 
electrons hopping coherently through the whole lattice. On the contrary, 
the strong correlation tries 
to optimize on-site (atomic) energetics, thus opposing against the hopping.   
This may involve two energy scales. The higher one is the so-called Hubbard $U$, 
which tends to suppress on-site valence fluctuations. The lower one, let us call it $J$,  
governs the splitting among on-site electronic configurations at fixed charge. 
It may be controlled by the exchange-splitting, the crystal field, 
by local distortion modes or even by short-range 
inter-site correlations. When the lattice model is driven towards a Mott metal-insulator 
transition (MIT), 
either by increasing $U$ or by doping at large $U$, it necessarily encounters 
a regime in which the coherent quasiparticle bandwidth $W_{qp}$ is of the same order as  
$J$, which we expect is essentially 
unaffected by $U$ as it just determines the multiplet splitting at fixed charge. 
Since coherent hopping tends to occupy more or less democratically all multiplets, 
it opposes against $J$. Out of this competition interesting physical properties 
may emerge, just like in the Anderson impurity models we discussed before. 
The analogy between the impurity and the lattice models 
can be even put on firm grounds in the limit of 
large coordination lattices through Dynamical Mean Field Theory (DMFT)\cite{DMFT}.    
In that limit it is possible to map the lattice model into an effective Anderson impurity 
model (AIM) subject to a self-consistency condition which relates the impurity Green's 
function to the hybridization with the conduction bath. The quasiparticle bandwidth 
of the lattice model transforms into the Kondo temperature $T_K$ of the AIM. 
Since approaching the MIT $W_{qp}\to 0$, the effective AIM is necessarily driven into 
the regime $T_K\sim J$, where the competition among the two screening mechanisms may 
result into anomalous physical properties. Exactly this competition was invoked 
by Ref.~\onlinecite{Capone-PRL} to explain the appearance of a superconducting 
pocket, later shown to have a hugely enhanced superconducting gap\cite{Capone-Science},  
just before the MIT in a model for alkali doped fullerenes.

More recently we have demonstrated by Wilson numerical renormalization group and 
by bosonization that a two-fold orbitally degenerate AIM in the presence of  
inverted Hund's rules possesses a non-Fermi liquid unstable fixed point similar to 
the two-impurity Kondo model one\cite{nostro}. 
Because of the aforementioned reasons, any lattice model which maps by DMFT 
into the same AIM should necessarily meet this fixed point on the route towards a MIT. 
We argued that, unlike the single-impurity, those lattice models might 
spontaneously generate by the DMFT self-consistency conditions  
a bulk order parameter to get rid of the single-impurity fixed-point 
instability. Since the fixed point is unstable in different, particle-hole and 
particle-particle, channels, there exist in principle several competing bulk 
instabilities. We speculated that, in the absence of nesting or band-structure 
singularities, the most likely instability is towards superconductivity. 
These predictions have been just recently confirmed on a lattice model 
by a DMFT calculation\cite{Capone-exE}. 
In this paper we pursue the analysis of that AIM by uncovering the spectral behavior 
across the non-Fermi liquid fixed point. This is not only interesting for the AIM itself, 
being one of the few cases where non-Fermi liquid dynamical properties may be accessed, 
but also in the context of the DMFT mapping. The model is also sufficiently simple to 
allow for an analytical description for the spectral function which reproduces  
well the numerical results and provide new physical insights.  
Actually our model spectral function has been quite useful in guiding the analysis of the 
DMFT solution presented in Ref.~\onlinecite{Capone-exE}.

The paper is organized as follows. In Section~\ref{The Model Hamiltonian} 
we describe the two-orbital AIM model. In Section~\ref{Physical Realizations} 
we introduce three lattice models which map by DMFT onto the two-orbital AIM: 
(a) a two-band Hubbard model $e\otimes E$ Jahn-Teller coupled to local phonons; 
(b) a two-band Hubbard model in the presence of single-ion anisotropy;  (c) 
two-coupled Hubbard planes. 
In Section~\ref{Numerical Renormalization Group Results} we review in more details 
the Wilson numerical renormalization group calculations of Ref.~\onlinecite{nostro} 
and present a new analysis based on Fermi liquid theory, which 
we develop in the Appendix. 
The new results concerning the dynamical properties are presented 
in Section~\ref{Spectral function}. 
In Section~\ref{Modelling} we extract from the numerical data 
an analytical expression of the impurity spectral function. The 
role of symmetry breaking terms in particle-hole channels is  
investigated  in Section~\ref{Particle-hole symmetry breaking terms}. 
Conclusions are presented in Section~\ref{Conclusions}.

\section{The Model Hamiltonian}
\label{The Model Hamiltonian}
The AIM Hamiltonian we consider is 
\begin{eqnarray}\label{Michele-ham}
  H & = & H_U + H_J + H_c + H_{hyb} \\
  & = & \frac{U}{2} \left(  n_{d} -2 + \nu\right)^2 + 2J \left[
  \left( T^x
  \right)^2 + \left( T^y \right)^2 \right] +\nonumber \\
  & & + \sum_{\vect{k}a\alpha} \epsilon_{\vect{k}} \, c^{\dagger}_{\vect{k} a
  \alpha} c^{\phantom{\dagger}}_{\vect{k} a \alpha} + \nonumber \\
  & & + \sum_{\vect{k}a\alpha} V_{d}
  \left( c^{\dagger}_{\vect{k} a \alpha} d^{\phantom{\dagger}}_{a\alpha} +
  d^{\dagger}_{a\alpha} c^{\phantom{\dagger}}_{\vect{k} a \alpha} \right).\nonumber 
\end{eqnarray}
Here $c^{\dagger}_{\vect{k} a \alpha}$ creates
a conduction electron in the 
band $a=1,2$ with momentum $\vect{k}$, spin $\alpha$ and 
energy $\epsilon_{\vect{k}}$, measured with respect to the 
chemical potential. $d^{\dagger}_{a\alpha}$ is the creation operator 
of an electron with spin $\alpha$ in 
the impurity orbital $a=1,2$, while $n_d = \sum_{a\alpha} d^\dagger_{a\alpha}
d^{\phantom{\dagger}}_{a\alpha}$ is the impurity occupation number. 
We have defined the orbital pseudo-spin operators 
\be
T^i = \frac{1}{2} \sum_\alpha \sum_{a=1,2} d^{\dagger}_{a\alpha}
\tau^i_{ab} d^{\phantom{\dagger}}_{b\alpha},
\label{T}
\ee
where $i=x,y,z$ and $\tau^i$'s are the Pauli matrices in the
orbital space. We further assume that the 
conduction band density of states is  
symmetric with respect to the chemical potential, set equal to zero, so that the 
behavior of the Hamiltonian under a particle-hole symmetry transformation is controlled 
by the parameter $\nu$ in (\ref{Michele-ham}). For the time being we 
will take $\nu=0$, which implies that the Hamiltonian is particle-hole symmetric. 
Afterwards we will release this constraint. 
The model without the impurity exchange coupling $J$ is 
SU(4) invariant. A finite $J$ lowers the SU(4) symmetry down to 
SU(2)$_{spin}\times$ O(2)$_{orbit}$. In this case the total charge, the total spin and the 
total $z$-component of the pseudospin are the only conserved quantities. 

It is convenient to start our analysis by the spectrum of the isolated impurity, $V_d=0$. 
The impurity eigenstates, $|n,S,S^z,T,T^z\rangle$, 
can be labeled by the occupation number $n$, the spin $S$, pseudospin $T$ and their 
$z$-components, $S^z$ and $T^z$, respectively, with energies 
\begin{equation}
E(n,S,S^z,T,T^z) = \frac{U}{2}(n-2)^2 + 2J\left[T(T+1)-\left(T^z\right)^2\right].
\label{eigenvalues}
\end{equation}
We assume $U\gg |J|$, so that the impurity ground state with $\nu=0$ has $n=2$. In this case 
the only configurations allowed by Pauli principle are a spin-triplet 
pseudo-spin-singlet, $S=1$ and $T=0$, 
\be
\begin{array}{lcl}
|2,1,+1,0,0\rangle &=& d^\dagger_{1\uparrow}d^\dagger_{2\uparrow}\,|0\rangle,\\
|2,1,0,0,0\rangle &=& \frac{1}{\sqrt{2}}
\left(d^\dagger_{1\uparrow}d^\dagger_{2\downarrow} 
- d^\dagger_{2\uparrow}d^\dagger_{1\downarrow}\right)
\,|0\rangle,\\
|2,1,-1,0,0\rangle &=& d^\dagger_{1\downarrow}d^\dagger_{2\downarrow}\,|0\rangle,\\
\end{array}
\label{triplet}
\ee
and a spin-singlet pseudo-spin-triplet, 
$S=0$ and $T=1$. The latter is split by $J$ into a singlet with $T^z=0$, 
\be
|2,0,0,1,0\rangle = \frac{1}{\sqrt{2}}
\left(d^\dagger_{1\uparrow}d^\dagger_{2\downarrow} 
+ d^\dagger_{2\uparrow}d^\dagger_{1\downarrow}\right)
\,|0\rangle,
\label{singlet}
\ee
and a doublet with $T^z=\pm 1$, 
\be 
\begin{array}{lcl}
|2,0,0,1,+1\rangle &=& d^\dagger_{1\uparrow}d^\dagger_{1\downarrow}\, |0\rangle, \\
|2,0,0,1,-1\rangle &=& d^\dagger_{2\uparrow}d^\dagger_{2\downarrow}\, |0\rangle. \\
\end{array}
\label{doublet}
\ee
If $J>0$, the lowest energy configuration is the spin-triplet, $S=1$ and $T=0$, 
which corresponds to the conventional Hund's rules. On the contrary, for 
$J<0$, the isolated impurity ground state is the singlet (\ref{singlet}) 
with quantum numbers $S=0$, $T=1$ and $T^z=0$. We postpone to the 
following Section a discussion about physical realization of such inverted Hund's rules.  
 
A finite hybridization, $V_d\not = 0 $, induces valence fluctuations within the 
impurity, which are controlled by the energy scale (hybridization width) 
\be
\Delta_0 = \pi \, V_d^2\, \rho_c,
\label{Delta}
\ee
with $\rho_c$ the conduction electron density of states (DOS) 
at the chemical potential per spin and band. 
These fluctuations are suppressed by    
a strong repulsion $U\gg \Delta_0$, which we assume throughout this work. 
Although all our calculations refer to the AIM (\ref{Michele-ham}), it is more insightful 
to discuss some physical properties in terms 
of the effective Kondo model which describes the low-energy behavior 
when $U\gg \Delta_0$:
\be
H_{eff} = H_J + H_c +H_K,
\label{Kondo-ham}
\ee
where $H_J$ and $H_c$ have been defined in (\ref{Michele-ham}) and the Kondo exchange 
\be
H_K = J_K\, \left[ \vec{S}\cdot\vec{{\cal S}} 
+ \vec{T}\cdot\vec{{\cal T}} + 4 \sum_{i,j=x,y,z} 
W_{ij}\, {\cal W}_{ij} \right],
\label{H_Kondo}
\ee
with 
\be
J_K = 2V_d^2/U.
\label{JKONDO}
\ee
Here $\vec{S}$, defined by 
\[
\vec{S} = \frac{1}{2}\sum_{a}\sum_{\alpha\beta} 
d^\dagger_{a\alpha}\, \vec{\sigma}_{\alpha\beta}\, 
d^{\phantom{\dagger}}_{a\beta},
\]
$\vec{T}$, which we introduced in Eq.~(\ref{T}),  
and $W_{ij}$,  
\[
W_{ij} = \frac{1}{4}\sum_{ab}\sum_{\alpha\beta} 
d^\dagger_{a\alpha}\, \tau^i_{ab}\,\sigma^j_{\alpha\beta}\, 
d^{\phantom{\dagger}}_{b\beta},
\]
are impurity spin, pseudo-spin and spin-orbital operators, respectively, while  
$\vec{{\cal S}}$, $\vec{{\cal T}}$ and ${\cal W}_{ij}$ are the corresponding conduction 
electron density operators at the impurity site. The impurity operators in 
(\ref{H_Kondo}) act only in the subspace with two electrons occupying the 
$d$-orbitals, which, as we showed, includes six states. 
The Kondo model (\ref{Kondo-ham}) contains two competing mechanisms 
which tend to freeze the left-over impurity degrees of freedom: ({\sl i}) the 
Kondo exchange, with its associated energy scale, the Kondo temperature $T_K$; 
({\sl ii}) the intra-impurity exchange splitting $J$. As we already mentioned, 
the Kondo exchange (\ref{H_Kondo}) gains energy by letting the impurity tunnel 
coherently among all available six configurations, but it is hampered by $J$ which instead 
tends to trap the impurity into a well defined state.   

If $J\gg T_K>0$, the positive exchange splitting dominates and the impurity is  
essentially frozen into the lowest energy spin-triplet configuration. 
The Kondo exchange projected onto the triplet sub-space (\ref{triplet}) is simply 
$H_K = J_K\, \vec{S}\cdot\vec{{\cal S}}$, describing a standard $S=1$ two-channel 
Kondo effect. This is known to be perfectly screened at low 
energy\cite{Nozieres&Blandin,Affleck&Ludwig}, 
yielding a scattering phase shift 
$\delta=\pi/2$ in each spin and orbital channel. 

On the contrary, if $J\ll -T_K<0$, the impurity gets trapped into the 
$S=0$, $T=1$ and $T^z=0$ configuration, Eq.~(\ref{singlet}). 
Since (\ref{singlet}) is non degenerate, the Kondo exchange is un-effective, so that 
asymptotically the impurity decouples from the conduction bath. This implies 
a low energy phase shift $\delta=0$. The main question which we try to adress is how 
the model moves across the two limiting cases. 

As it was pointed out in Ref.~\onlinecite{nostro}, this behavior is parallel 
to the two $S=1/2$ impurity Kondo model (2IKM) in the presence of a 
direct exchange between the 
impurity spins\cite{J&V,A&L,AL&J}. In that case, if the two spins are strongly 
ferromagnetically coupled, the model reduces to an $S=1$ two-channel Kondo model, 
while, if they are strongly antiferromagnetically coupled, the two spins bind  
together into a singlet and decouple from the conduction electrons, exactly as in our model. 
The two channels correspond in the 2IKM to the symmetric and antisymmetric 
combinations of the even and odd scattering channels with respect to the midpoint 
between the impurities.  
It was demonstrated by Ref.~\onlinecite{AL&J} that, provided a peculiar particle-hole symmetry 
holds, the non-Fermi liquid unstable fixed point (UFP) 
found by Ref.~\onlinecite{J&V} separates the Kondo screened 
and unscreened regimes. In particular it was shown that while a particle-hole 
symmetry breaking term
\be
\delta H_{p-h} = -\mu_d \sum_{a\alpha}\, d^\dagger_{a\alpha}d^{\phantom{\dagger}}_{a\alpha} 
- \sum_{\vect{k},a\alpha}\, \mu_{\vect{k}}\, c^\dagger_{\vect{k}a\alpha}
c^\dagga_{\vect{k}a\alpha},
\label{p-h}
\ee
does not wash out the UFP, the latter is instead  destabilized by the perturbation 
\bea 
\delta H_{rel} = - h_d\, \sum_\alpha d^\dagger_{1\alpha}d^\dagga_{2\alpha} + H.c.  \nonumber \\
- \sum_{\vect{k},\alpha}\, h_{\vect{k}}\, c^\dagger_{\vect{k}1\alpha}
c^\dagga_{\vect{k}2\alpha} + H.c.\, .
\label{rel}
\eea
Translated into our two-orbital language, the dangerous symmetry 
which needs to be preserved is just the O(2)$_{orbit}$ orbital symmetry. 
Therefore, unlike in the 2IKM, where the 
two scattering channels are generically not degenerate, in our case the 
instability towards O(2)$_{orbit}$ symmetry breaking does correspond to a physical 
instability. Hence, if orbital symmetry is unbroken, we do expect to find an UFP in our 
model, with similar properties as in the 2IKM. We notice that, in spite of the analogies, 
our model has a larger impurity Hilbert space than the 2IKM. In fact the 
$S=0$, $T=1$ and $T^z=\pm 1$ doublet of Eq.~(\ref{doublet}) is absent in the 2IKM, where 
it would correspond to doubly occupied impurities (the labels 1 and 2 for the $d$-orbitals 
translate in the 2IKM into the two one-orbital impurities). Yet we can perturb our 
Hamiltonian by adding to $H$ of (\ref{Michele-ham}) the term
\be
H_G = G\, \left(T^z\right)^2,
\label{towards2IKM}
\ee
with $G>0$, which raises the energy of the doublet. If $G\gg T_K$, the doublet 
effectively decouples from the low energy sector, and our model should become 
equivalent to the 2IKM. 
In Section~\ref{Numerical Renormalization Group Results}  we show that indeed 
by increasing $G$ our UFP smoothly transforms into the 2IKM one.  

\section{Physical Realizations}
\label{Physical Realizations}
As we emphasized in the Introduction, our interest in model (\ref{Michele-ham}) plus eventually 
(\ref{towards2IKM}) is mainly motivated by its possible relevance for  
lattice models. In reality a formal correspondence 
bewteen single-impurity and lattice models holds strictly only 
in the limit of large lattice-coordination. Nevertheless we believe that 
this correspondence, at least close to a Mott transition, may remain 
valid even beyond that limit, making the single-impurity analysis of much broader interest. 
Therefore, although inversion of Hund's rules may 
indeed occur in realistic AIM's or in artificially designed quantum dot devices, here we rather 
focus on lattice models which map within DMFT into our AIM.

\subsection{Two-band Hubbard model in the presence of an $e\otimes E$ Jahn-Teller coupling}

Let us start by considering a two-band Hubbard model in which each site 
is Jahn-Teller coupled to a doubly degenerate phonon. The Hamiltonian reads 
\bea
H &=&  -\frac{t}{\sqrt{z}}\sum_{a=1}^2\sum_{\sigma}\sum_{<ij>} \, 
\left(c^\dagger_{ai\sigma}c^\dagga_{aj\sigma} + H.c.\right)\nonumber \\
&& + \frac{U}{2}\sum_i (n_i-2)^2 + 
2J_H\sum_i \left[ \left(T^x_i\right)^2 + \left(T^y_i\right)^2 \right]\nonumber \\
&& + \frac{\omega_0}{2}\sum_i \sum_{a=x,y} \left(q_{ia}^2 + p_{ia}^2\right) 
- g\sum_i \left(q_{ix}\, T^x_i + q_{iy}\, T^y_i\right).
\label{HJT}
\eea   
Here $-t/\sqrt{z}$ is the hopping matrix element bewteen one site and its $z$-neighbors and 
$J_H>0$ is a conventional Hund's exchange.  
$q_{ix}$ and $q_{iy}$ are the phonon coordinates at site $i$, $p_{ix}$ and $p_{iy}$ 
their conjugate 
momenta, $\omega_0$ the phonon frequency and $g$ the Jahn-Teller coupling. 
The latter gives rise to 
a retarded electron-electron interaction whose Fourier transform is 
\[
g^2 \sum_i \, \frac{\omega_0}{\omega^2-\omega_0^2}\left[ T^x_i(\omega)\, T^x_i(-\omega) + 
T^y_i(\omega)\, T^y_i(-\omega)\right].
\]
If the phonon frequency $\omega_0$ is much larger than the quasiparticle 
bandwidth we can safely neglect the $\omega$-dependence at low energy,  
so that the phonon-mediated interaction becomes un-retarded and given by 
\[
-\frac{g^2}{\omega_0}\sum_i \, \left[ \left(T^x_i\right)^2 + \left(T^y_i\right)^2\right].
\]
Within DMFT the Hamiltonian maps in the large $z$-limit onto the same AIM model as in 
Eq.~(\ref{Michele-ham}) with 
\be
J = J_H -\frac{g^2}{2\omega_0},
\label{mapJT}
\ee
which may be either positive or negative. The case with $J<0$ 
as well as the starting model realistically including 
phonons have been recently studied 
by DMFT\cite{Capone-exE,Han}.

\subsection{Two-band Hubbard model with single-ion anisotropy}

Another realization which may also be physically relevant is the 
following lattice model:
\bea
H &=&  -\frac{t}{\sqrt{z}}\sum_{a=1}^2\sum_{\sigma}\sum_{<ij>} \, 
\left( c^\dagger_{ai\sigma}c^\dagga_{aj\sigma} + H.c.\right) \nonumber \\
&& + \frac{U}{2}\sum_i (n_i-2)^2 - 
2J_H\sum_i \vec{S}_i\cdot\vec{S}_i \nonumber \\
&& + D\sum_i \left(S^z_i\right)^2.
\label{H_Hund}
\eea   
For $J_H>0$ and $D\not=0$ this model describes a two-band Hubbard model with conventional 
Hund's rules, favoring a spin-triplet two-electron configuration, in the presence 
of a single-ion anisotropy which splits the spin-triplet into a singlet with $S^z=0$ 
and a doublet with $S^z=\pm 1$. If $D>0$, the $S^z=0$ configuration is favored. 
Upon interchanging $\vec{S} \leftrightarrow \vec{T}$, 
this model maps in the $z\to\infty$ limit onto (\ref{Michele-ham}) with 
\be
\begin{array}{lcl}
J &=& -J_H, \\
G &=& D-2J_H.\\
\end{array}
\label{map-H_Hund}
\ee

\subsection{Two coupled Hubbard planes}
Finally let us consider two coupled single-band Hubbard planes described by the Hamiltonian
\bea
H &=& -\frac{t}{\sqrt{z}}\sum_{a=1}^2\sum_{\sigma}\sum_{<ij>} \,  
\left(c^\dagger_{ai\sigma}c^\dagga_{aj\sigma}+H.c.\right) 
+ \frac{U}{2}\sum_{a,i} \, \left(n_{ai}-1\right)^2 \nonumber\\ 
&& + \sum_i \, J\,\vec{S}_{1i}\cdot\vec{S}_{2i}\, +\, V\, (n_{1i}-1)(n_{2i}-1),
\label{H-Hubbard}
\eea
where $a=1,2$ labels the two planes and $-t/\sqrt{z}$ is the 
in-plane hopping between one site and its 
$z$-neighbors. 
In the limit $z\to\infty$, this model maps by DMFT onto an 
AIM self-consistently coupled to a bath\cite{DMFT}. 
The relations between the interaction parameters of the 
AIM, (\ref{Michele-ham}) plus (\ref{towards2IKM}),  
and those of (\ref{H-Hubbard}) are given in Table~\ref{map-Hubbard}.
\begin{table}[h]
\caption{Mapping between the AIM interaction parameters and the 
two Hubbard plane ones. }
\label{map-Hubbard}
   \begin{tabular}{||c|c||}
      \hline\hline
AIM & Two Hubbard planes \\
\hline \hline
$U$ & $\frac{1}{2}\left(U+V\right) - \frac{1}{8}J$\\ \hline
$J$ & $-\frac{1}{4} J$ \\ \hline
$G$ & $U-V +\frac{1}{4} J$\\ \hline \hline
   \end{tabular}
\end{table}

In reality it is more interesting to consider the model (\ref{H-Hubbard}) with 
$J=V=0$ but in the presence of an inter-plane  hopping
\be
-t_\perp \sum_{i\sigma}\left( 
c^\dagger_{1i\sigma}c^\dagga_{2i\sigma} + H.c.\right).
\label{tperp}
\ee
In the limit of large lattice-coordination, this model maps close to the MIT 
onto a two-orbital AIM with an hybridization width at the chemical potential 
much smaller than $U$. Since by the Table~\ref{map-Hubbard} $G=U$, we can 
safely project out of the low energy subspace the doublet (\ref{doublet}). 
The effective AIM within the impurity subspace which includes the singlet 
(\ref{singlet}) and the spin-triplet is 
\bea
H_{AIM} &=& H_c \,+\, J_K\, \left(\vec{S}_1+\vec{S}_2\right)\cdot\vec{\cal{S}} \nonumber \\
&& + \,J\, \vec{S}_1\cdot\vec{S}_2 \,+\, 
J_K\, \frac{t_\perp}{U}\, {\cal{T}}^x,
\label{2HP-2IKM}
\eea
where $H_c$ and $J_K$ have been defined in Eqs.~(\ref{Michele-ham}) and (\ref{JKONDO}), 
$\vec{S}_1$ and $\vec{S}_2$ are the impurity spin operators for 
the singly-occupied orbitals 1 and 2, 
while $\vec{\cal{S}}$ and ${\cal{T}}^x$ are respectively the 
conduction-electron 
spin-density operator and $x$-component 
of the pseudo-spin density operator, $\vec{\cal{T}}$, at the impurity site.  
The impurity antiferromagnetic exchange, $J=4t_\perp^2/U$, 
lowers the energy of the 
singlet (\ref{singlet}) with respect to the spin triplet. 
Therefore $J$ alone might induce an UFP within the phase diagram, 
just like in our model as well as in the 2IKM. 
However $t_\perp$ also introduces a ${\cal{T}}^x$   
scattering potential at the impurity site, last term in the right hand side 
of Eq.~(\ref{2HP-2IKM}), which is known to be relevant at the UFP.  
In this respect $t_\perp$ plays an intriguing role: 
on one hand it provides a mechanism, the 
antiferromagnetic exchange $J$, able to stabilize a non-trivial fixed point, 
but, in the meantime,  it also prevents that fixed point to be ever reachable. 
Yet we might wonder whether the critical region around the UFP is completely 
or only partially washed out. In the latter case we should expect that 
the physics of the lattice model close to the MIT is still influenced  
by the UFP, with interesting consequences. We will come back to this issue 
in Section~\ref{Particle-hole symmetry breaking terms}.

\section{Numerical Renormalization Group Results} 
\label{Numerical Renormalization Group Results} 
To study the AIM (\ref{Michele-ham}) we used the Wilson Numerical Renormalization Group
(NRG) method\cite{NRG}. This technique is known to provide a detailed
description of the low energy behavior thus allowing a faithful 
characterization of the fixed points as well as of their stability domain. 
In addition dynamical properties are also accessible by NRG, which, as we are going to show, 
are of notable interest. 

Within NRG the conduction band is logarithmically discretized and mapped 
onto a one-dimensional chain with  
nearest neighbor hopping integrals which decrease exponentially along the
chain. The Hamiltonian of a chain with $N+1$ sites is defined by: 
\begin{eqnarray}\label{N-form}
  H_N & = & \Lambda^{(N-1)/2} \Bigg\{ \sum_{n=0}^{N-1} \Lambda^{-n/2}
  \xi_n \left( c^\dagger_{na\alpha} c^\dagga_{(n+1)a\alpha} + c^\dagger_{(n+1)a\alpha}
  c^\dagga_{na\alpha} \right) + \nonumber \\
  &&+ \tilde{\Delta}_0^{1/2} \left( c^\dagger_{0a\alpha} d^\dagga_{a\alpha} +
  d^\dagger_{a\alpha} c^\dagga_{0a\alpha} \right) +\nonumber \\
  &&+ \frac{\tilde{U}}{2} \left(  n_{d} -2 \right)^2 + 2\tilde{J} \left[
  \left( T^x
  \right)^2 + \left( T^y \right)^2 \right] \Bigg\}.
\end{eqnarray}
Here $\tilde{U} = C_\Lambda \, U $,
$\tilde{J} = C_\Lambda \, J$,
$\tilde{\Delta}_0 = C_\Lambda^2 \, \frac{2\Delta_0}{\pi }$, see (\ref{Delta}), 
where $C_\Lambda = \left( \frac{2\Lambda}{1+\Lambda} \right)$ and all energies 
are measured in units of half the conduction bandwidth. 
The re-scaling factor $\Lambda^{(N-1)/2}$ in front of (\ref{N-form}) 
keeps the lowest energy scale in $H_N$ of order one at each iteration. 
The original Hamiltonian is recovered in the limit of infinite chain lenght:
\begin{equation}
   H  = \lim_{N \rightarrow \infty} C_\Lambda^{-1}
   \Lambda^{-(N-1)/2} H_N. 
\end{equation}
The size $N$ of the chain determines the infrared cut-off, {\sl e.g} 
the temperature $T$, on a logarithmic scale ($T \sim \Lambda^{-N/2}$).
The method essentially consists in diagonalizing the model on a finite size chain, 
{\sl e.g.} $N$,   
and iteratively increasing the size by one site, from $N$ to $N+1$, keeping 
only the low energy part of the $N$-site spectrum. (In our calculations 
we typically kept up to the lowest 2000 states, not counting degeneracies, 
and used $\Lambda$ between 2 and 3. We did  
check that these numbers are sufficient to get accurate results.)   

We restrict our analysis to large values of $U$, where valence fluctuations 
on the impurity are substantially suppressed. Here, as we discussed, the AIM 
effectively behaves like the Kondo model (\ref{Kondo-ham}).
We fix both $U$ and
$\Delta$ and span the phase space by varying the exchange parameter $J$. 

\subsection{Low energy spectrum at the fixed points}
\begin{figure}
\centerline{
\includegraphics[width=7.5cm]{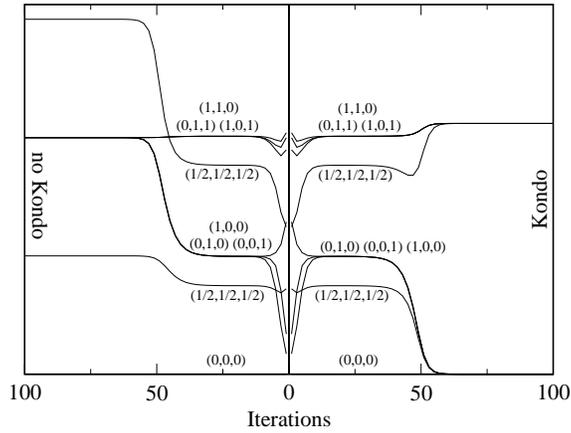} }
\caption{Lowest energy levels versus the chain size $N$. The right/left panels 
correspond to a deviation $\delta J/J^* = \pm 4\cdot 10^{-3}$ from
the fixed point value $J^*$. The levels are labeled by the quantum numbers $(Q,T^z,S)$, 
where $Q$ is one-half of the added charge with respect to the ground state value.  
\label{spettro}
}
\end{figure}

First of all we identify the fixed points by analysing the low energy spectra 
(with $N$ typically up to one hundred). 
Since the conventional size-dependence of the level spacing 
is absorbed by the factor $\Lambda^{(N-1)/2}$ in front of (\ref{N-form}), 
the low-lying energy levels flow to constant values whenever the model is close to  
a fixed point. Fig.~\ref{spettro} shows that there are two different asymptotic 
regimes separated by a critical value $J^*<0$. 
In order to facilitate the interpretation of that figure, we recall 
that the ground state of a  
particle-hole symmetric free-chain with 
$N+1$ sites is unique if $N$ is odd and degenerate if $N$ is even. 

For $J > J^*$ the low energy spectrum of a
chain with {\em odd} number, $N+1$, of sites flows towards that of a free chain  
with an {\em even} number of sites and viceversa. This is evident  
in the right panel of Fig.~\ref{spettro} where the ground state of 
the chain with odd $N$ becomes asymptotically degenerate as for a chain with even $N$. 
Apart from the ground state degeneracy, also the low-lying spectrum, 
{\i.e.} degeneracy and quantum numbers of the levels as well as the level spacings, 
coincides with that of a free chain.   
As usual, this is as if the first site of the chain were locked to
form a spin and orbital singlet configuration with the impurity, hence 
becoming unaccessible to the conduction electrons which thus 
acquire a $\pi/2$ phase shift per conduction channel. It is a 
conventional Kondo screened phase.

For $J < J^*$ the situation is reversed: the low energy spectrum
of an odd (even) chain flows to that of an 
odd (even) free chain. Indeed, as shown in the left panel of Fig.~\ref{spettro}, 
the ground state with $N$ odd remains non-degenerate for large $N$.
This case corresponds to an unscreened phase with the impurity 
asymptotically decoupled from the conduction bath. 
The phase shift is consequently $\delta=0$.

In between the Kondo screened and unscreened phases we do find a non-trivial fixed point, 
as it is visible in the intermediate cross-over region of the spectrum, 
see Fig.~\ref{spettro}.  
The peculiar non-Fermi liquid character of this intermediate coupling 
unstable fixed point (UFP) is clear by the non-uniform
spacing of the low energy levels. A careful analysis of the UFP spectrum  
reveals that it is just the same as that one found in the particle-hole-symmetric 
2IKM\cite{AL&J}. 
In Table~\ref{table} we compare the energies $E$ of the lowest-lying levels of 
the Wilson chain at the UFP,  
as obtained by NRG, with the prediction $x$ of Conformal Field Theory for the 2IKM\cite{AL&J}.  
The agreement is a clear evidence that the UFP is indeed the same in both models.

\subsection{Impurity properties at the UFP}
\begin{figure}
\centerline{
\includegraphics[width=7.5cm]{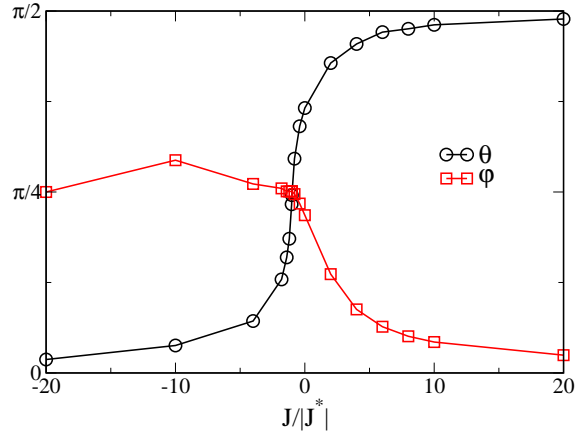} }
\caption{The angles $\theta$ and $\phi$ as defined 
through Eq.~(\ref{Prob-vs-av}). Notice that the fixed point is identified 
by $\theta=\phi=\pi/4$.
\label{theta-phi}
}
\end{figure}

\begin{figure}
\centerline{
\includegraphics[width=7.5cm]{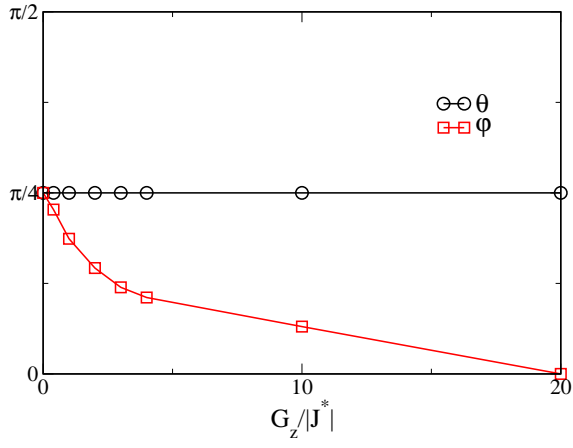} }
\caption{The UFP values of $\theta$ and $\phi$ along the path 
parametrized by the coupling $G$ from our 
to the 2IKM model.  
\label{av-val-G}
}
\end{figure}

Additional information are provided by the average values of the 
impurity spin, $\langle \vec{S}\cdot\vec{S}\rangle$, psudospin, 
$\langle \vec{T}\cdot\vec{T}\rangle$, and its $z$-component, 
$\langle \left(T^z\right)^2\rangle$. By symmetry, the impurity density matrix 
is diagonal in the six two-electron configurations. 
The diagonal elements represent the occupation probabilities $P(S,S^z,T,T^z)$ 
of states with quantum numbers $S$, $S^z$, $T$ and $T^z$. In the 
large $U$-limit, where impurity configurations with $n\not =2$ have negligible weight, 
we can write  
\be
\begin{array}{lcl}
P(0,0,1,0) &=& \cos^2 \theta,\\
P(0,0,1,+1) &=& P(0,0,1,-1) = \frac{1}{2}\,\sin^2\theta\, \sin^2\phi, \\
P(1,+1,0,0) &=& P(1,0,0,0)=P(1,-1,0,0)=\frac{1}{3}\, \sin^2\theta \, \cos^2\phi,\\
\end{array}
\label{Prob}
\ee
from which it derives that 
\be
\begin{array}{lcl}
\langle \vec{S}\cdot\vec{S}\rangle &=& 2\, \sin^2\theta \, \cos^2\phi,\\
\langle \vec{T}\cdot\vec{T}\rangle &=& 2\,
\left(\cos^2 \theta + \sin^2\theta\, \sin^2\phi\right), \\
\langle \left(T^z\right)^2\rangle  &=& \sin^2\theta\, \sin^2\phi.\\
\end{array}
\label{Prob-vs-av}
\ee
In Fig.~\ref{theta-phi} we plot the angles $\theta$ and $\phi$ as obtained through 
(\ref{Prob-vs-av}) by the numerical calculated average values. 
The UFP is characterized by $\theta=\phi=\pi/4$, namely by the value 1/2 of the 
occupation probability of the singlet state (\ref{singlet}). The precise value of 
the other occupation probabilities, in other words of $\phi$, are instead not relevant, 
apart from the obvious 
fact that their sum should be 1/2 too. In fact, if we add the term (\ref{towards2IKM}) 
with $G>0$, we do find the same UFP, which locations now depends also on $G$, 
which is still identified by $P(0,0,1,0)=1/2$, {\sl i.e.} $\theta=\pi/4$, although the 
weight of the spin-triplet is enhanced with respect to the doublet (\ref{doublet}), 
$\phi<\pi/4$. 
For large $G$ we do recover the 2IKM values $\theta=\pi/4$ and $\phi=0$, 
see Fig.~\ref{av-val-G}.

\subsection{Approach to the fixed points}
     
As we said the low energy spectrum both in the Kondo screened and unscreened phases 
flows to that of a free chain, with one less site in the former case. The flow towards 
the asymptotic spectrum can be described by a free chain in the presence 
of a local perturbation term\cite{NRG} acting on the 
first available site, denoted as site $0$, of the conduction chain, 
which is actually the second site in the 
Kondo screened phase. By symmetry considerations this local term can be 
in general written as: 
\bea
\delta H_* &=& -t_*\sum_{a\alpha} \left(c^\dagger_{0a\alpha}c^\dagga_{1a\alpha} 
+ H.c.\right) + \frac{U_*}{2}\left(n_0-2\right)^2 \nonumber \\
&& + J_{S*}\, \vec{\cal{S}}_0\cdot\vec{\cal{S}}_0 
+ J_{T*}\, \vec{\cal{T}}_0\cdot\vec{\cal{T}}_0 
-2\left(J_{S*}+J_{T*}\right)\, \left({\cal{T}}_0^z\right)^2.
\label{Hfp}
\eea
We choose this particular form because it has the advantage that the energy of the 
center of gravity of each multiplet with given charge $n_0$ is just $U_*(n_0-2)^2/2$.
Upon approaching the UFP on both sides, we find that $U_*\sim 
J_{S*}=\gamma\rightarrow +\infty$, 
$J_{T*}\sim -5\gamma \rightarrow -\infty$ and $t_*\sim 3\gamma/8\to +\infty$.
The behavior of $t_*$ implies a divergence of the impurity contribution to the 
specific heat coefficient. Namely if $\delta C_V$ is the variation of the 
specific heat with respect to its value $C_V$ in the absence of the impurity, then 
\[
\frac{\delta C_V}{C_V} \sim \rho_c\, t_* \to \infty.
\]

\begin{figure}
\centerline{
\includegraphics[width=8cm]{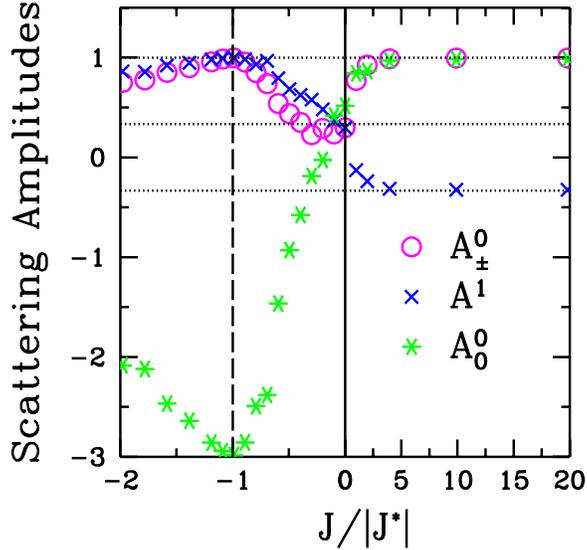} }
\caption{The scattering amplitudes 
in the various particle-particle channels as function of $J$ measured in units of the 
UFP $J^*$. Notice the agreement with the values 
predicted by general arguments 
presented in the Appendix at the UFP, $J/|J^*|=-1$, 
at the SU(4) point, $J/|J^*|=0$, and in the limit of the $S=1$ two-channel Kondo model, 
$J/|J^*|\gg 1$.  
\label{FL:gamma}
}
\end{figure}

In reality, it is more convenient to analyse the NRG results by invoking 
the Fermi liquid theory which we present in the Appendix. Through NRG, 
one can calculate the Wilson ratios related to the conserved quantities\cite{NRG}, namely the 
total charge, spin and $z$-component of $\vec{T}$. If $\delta \chi_i$ is  
the variation of the susceptibility with respect to the value $\chi_i$ without 
the impurity, where $\chi_i=\chi_C,\chi_S,\chi^{||}_T$ are the 
charge, spin, and $T^z$ susceptibilities, then 
the Wilson ratios $R_i$ are defined through 
\[
R_i = \frac{\delta \chi_i}{\chi_i}\, \frac{C_V}{\delta C_V}.
\]
On the other hand, Fermi liquid theory implies also that 
\be
R_i = 1 - A_i,
\label{Wilson-ratios}
\ee
where $A_i$ is the dimensionless quasi-particle scattering amplitude 
in channel $i$ defined in Eq.~(\ref{FL:def:A}) through the scattering vertex at low  
incoming and outgoing frequencies and the quasi-particle density of states at the 
chemical potential, see (\ref{FL:def:z}). In general we can introduce a scattering 
amplitude for each particle-hole and particle-particle channel. In particular, 
besides $A_C$, $A_S$ and $A^{||}_T$, we consider the particle-hole scattering amplitudes 
in the $T^x$ channel, which is degenerate with the $T^y$ one, $A^{\perp}_T$, 
as well as in the spin orbital channels $\vec{S}\, T^z$, $A^{||}_{ST}$, and  
$\vec{S}\, T^{x(y)}$, $A^{\perp}_{ST}$. In addition we introduce the 
amplitudes in the particle-particle channels, namely $\A^1$ in the 
spin-triplet orbital-singlet Cooper channel, $\A^0_0$ and $\A^0_{\pm}$ in the spin-singlet 
orbital-triplet channels with $T^z=0$ and $T^z=\pm 1$, respectively. As shown in the 
Appendix, all particle-hole scattering amplitudes can be expressed through the 
particle-particle ones: 
\begin{eqnarray}
A_C &=& \frac{1}{4}\left(6\A^1 + 2\A^0_0 + 4\A^0_{\pm}\right),
\label{NRG:Ac}\\
A_S &=& \frac{1}{4}\left(2\A^1 - 2\A^0_0 - 4\A^0_{\pm}\right),
\label{NRG:AS}\\
A^{||}_T &=& \frac{1}{4}\left(-6\A^1 - 2\A^0_0 + 4\A^0_{\pm}\right),
\label{NRG:ATz}\\
A^{\perp}_T &=& \frac{1}{4}\left(-6\A^1 + 2\A^0_0 \right),
\label{NRG:ATxy}\\
A^{||}_{ST} &=& \frac{1}{4}\left(-2\A^1 + 2\A^0_0 - 4\A^0_{\pm}\right),
\label{NRG:ASTz}\\
A^{\perp}_{ST} &=& \frac{1}{4}\left(-2\A^1 - 2\A^0_0 \right).
\label{NRG:ASTxy}
\end{eqnarray}
Since we are able to calculate by NRG the three Wilson ratios $R_C$, which is zero 
in the Kondo limit, $R_S$ and $R^{||}_T$, we can also determine the three unknown 
particle-particle scattering amplitudes through 
Eqs.~(\ref{Wilson-ratios})-(\ref{NRG:ATz}), which we plot 
in Fig.~\ref{FL:gamma}.

The first thing to notice is that approaching the UFP, 
\[
\A^0_0\simeq A^{\perp}_T\simeq A^{||}_{ST}\simeq -3,
\]
while all the other $A_i$'s tend to 1, implying vanishing Wilson ratios. 
The fixed point seems therefore to display 
a large hidden symmetry, actually an SO(7) as identified by Ref.~\onlinecite{AL&J}. 
The UFP is equally 
unstable in the $s$-wave Cooper channel with $S=0$, $T=1$ and $T^z=0$,  
as well as in the particle-hole $T^{x(y)}$ and $\vec{S}\, T^z$ channels\cite{nostro,nota}. 
All of them correspond to physical instabilities as 
we anticipated and unlike what happens in the 2IKM. 
On the contrary any external field in the other channels do not spoil 
the UFP, in particular in the charge, spin and $T^z$ particle-hole channels, 
which refer to conserved quantities.

The physics around and right at the UFP has been uncovered by 
Conformal Field Theory and bosonization\cite{A&L,AL&J,Varma,Gan,nostro}. 
Due to the existence of 
two energy scales, the Kondo temperature $T_K$ and the exchange splitting, $J$, 
the quenching of the impurity degrees of freedom takes place in two steps. 
First, around an energy scale $T_+\sim \max(T_K,|J|)$, most of the $\ln 6$ entropy of the 
two-electron impurity multiplets is removed, leaving behind a residual entropy 
$\ln\sqrt{2}$ which gets quenched only below a lower energy scale $T_-\sim 1/\gamma$.   
The latter depends quadratically upon the deviation from the UFP, 
namely $T_-\sim |J-J^*|^2/T_+$. 
The entropy has a low energy linear behavior, $S(T)\sim T/T_-$, followed 
above $T_-$ by another linear one, $S(T)-\ln\sqrt{2}\sim T/T_+$\cite{AL&J,Gan}. 
At the fixed point, $T_-=0$, 
there is a finite residual entropy $S(0)=\ln\sqrt{2}$ and $S(T)-S(0)\sim T/T_+$. 
A perturbation in any of the relevant channels washes out the fixed point 
cutting off the infrared singularities close to 
the UFP on an energy scale which depends quadratically upon the strength of the 
perturbation. 
In Section~\ref{Particle-hole symmetry breaking terms} 
we analyse more explicitly the stability/instability  
of the UFP towards symmetry breaking fields in particle-hole channels. 

\subsection{Influences of the single-impurity behavior in a DMFT calculation}
Let us now instead discuss the above results in connection with DMFT. Suppose there is 
a lattice model which maps in the limit of large lattice-coordination onto 
the AIM (\ref{Michele-ham}) with $J<0$. If the model is driven towards 
a Mott metal-insulator transition, the effective AIM is necessarily pushed 
into a regime in which $T_K\sim |J|$, namely in the critical region around the 
UFP. As shown in Fig.~\ref{FL:gamma}, the $s$-wave scattering amplitude 
$\A^0_0$ as well as the equally relevant $A^{\perp}_T$ and $A^{||}_{ST}$ 
are strongly attractive in an entire interval 
around the UFP. This suggests that the impurity 
fixed-point instability might transform by DMFT self-consistency into a   
whole pocket where the model generates spontaneously a bulk symmetry-breaking 
order-parameter along one of the relevant channels. As we argued in 
Ref.~\onlinecite{nostro}, if nesting or Van Hove singularities are absent,  
it is most probable that the dominant instability will occur in the Cooper channel, 
the only one which is singular in any dimensions and for any band-structure 
with a finite quasi-particle density of states at the chemical potential. This has been indeed 
confirmed by very recent DMFT calculations in Refs.~\onlinecite{Capone-exE,Han}. 

The other interesting observation is that in the conventional Hund's regime, 
the Kondo screened phase with $J>0$, an attraction in the spin-triplet 
$T=0$ channel develops, $\A^1<0$. In realistic lattice models which map onto the 
AIM with $J>0$ in 
the limit of large lattice-coordination, spin-triplet superconductivity would compete 
with bulk magnetism. 
Yet, if magnetism is frustrated, spin-triplet superconductivity might emerge. 
In particular, since increasing the Hubbard $U$ in the lattice model implies 
decreasing $T_K$ in the AIM, which is the same as increasing the 
effective strength of $J>0$, we should expect that spin-triplet superconductivity 
is enhanced near the MIT. This has been recently observed by  
DMFT\cite{Han}. However the enhancement of the spin-triplet amplitude is 
not as dramatic as for the spin-singlet one near the UFP at $J<0$. 
This situation would change in the presence of a single-ion 
anisotropy which favors e.g. spin-triplet pairing with $S^z=0$, 
see the model Eq.~(\ref{H_Hund}). As we showed, this model 
is equivalent to (\ref{Michele-ham}) upon interchanging 
the role of $\vec{T}$ with $\vec{S}$. This suggest that the lattice model 
which maps by DMFT onto (\ref{H_Hund}) with $D>0$ would still enter a local critical regime 
before the MIT. Here it should be dramatically enhanced 
the tendency towards spontaneous generation of a bulk order parameter 
in the particle-hole channels $S^x$, $S^y$ and $\vec{T}\,S^z$ 
as well as in the spin-triplet Cooper channel 
with $S^z=0$: 
$
\left(c^\dagger_{1\uparrow}\,c^\dagger_{2\downarrow} 
- c^\dagger_{2\uparrow}\,c^\dagger_{1\downarrow}\right)
$.

\section{Impurity spectral function}
\label{Spectral function}

The impurity density of states (DOS), $\rho(\epsilon)$, is defined through 
\be
\rho(\epsilon) = -\frac{1}{2\pi}\, \lim_{\eta\to 0}  
\left[G(\epsilon+i\eta)-G(\epsilon-i\eta)\right],
\ee
where $G(i\epsilon_n)$ is the impurity Green's function in Matsubara frequencies, 
which, by symmetry, is diagonal in spin and orbital indices, and independent upon them. 
In general 
\be
G(i\epsilon_n)^{-1} = i\epsilon_n - \Delta(i\epsilon_n) - \Sigma(i\epsilon_n) 
= G_0(i\epsilon_n)^{-1} - \Sigma(i\epsilon_n),
\label{ISF:Green}
\ee
where $G_0(i\epsilon_n)$ is the non-interacting, $U=J=0$, Green's function,  
\be
\Delta(i\epsilon_n) = V_d^2\, \sum_{\vect{k}} \frac{1}{i\epsilon_n - \epsilon_{\vect{k}}},
\label{ISF:hybr}
\ee
is the hybridization function, and $\Sigma(i\epsilon_n)$ the impurity self-energy. 
Let us suppose to follow the behavior of the DOS as the interaction is switched on. 
We will imagine to increase slowly both $U$ and $|J|$ at fixed $U/|J|\gg 1$ with $J<0$. 
When $U$ is small, one can show by perturbation theory that  
\[
{\cal I}m \Sigma(\epsilon) \sim \epsilon^2, 
\]
which is the standard result that the quasiparticle decay rate vanishes faster than the 
frequency. Therefore at the chemical potential, $\epsilon=0$, the 
impurity DOS is not affected by a weak interaction, since   
\be
\rho(0) = -\frac{1}{\pi}\, \lim_{\eta\to 0} {\cal I}m \, G(0+i\eta) 
= -\frac{1}{\pi}\, \lim_{\eta\to 0} {\cal I}m \, G_0(0+i\eta) = 
\frac{1}{\pi\Delta_0}=\rho_0, 
\ee  
where $\Delta_0 = -{\cal I}m \, \Delta(0+i\eta)$ was introduced in Eq.~(\ref{Delta}), 
and $\rho_0$ denotes the non-interacting DOS at the chemical potential.    
In a single-orbital AIM, the above result remains valid even when the interaction 
is very large. In our case we 
may expect that something non-trivial should instead occur. Indeed, upon increasing 
$U$, the AIM enters the Kondo regime, with a Kondo temperature exponentially decreasing 
with $U$. Therefore at some critical $U_c$, when $T_K\sim |J|$, the AIM has to cross the 
non-Fermi liquid UFP. Namely the UFP of our AIM 
can also be attained by increasing the interaction strength, signaling a breakdown 
of the conventional perturbation theory.  
We now discuss how this criticality shows up in the 
spectral properties.   

The impurity DOS can be obtained by NRG by directly evaluating the spectral function  
\begin{equation}
   A_{a\alpha}(\omega) = \frac{1}{Z}\sum_{m,n} |\langle m|d^\dagger_{a\alpha}
   | n \rangle|^2 \delta \left(\omega - (E_n -E_m)\right)
   \left( e^{-\beta E_n} + e^{-\beta E_m} \right).
\end{equation}

For any finite chain $A(\omega)$ is a discrete sum of delta-peaks.  
A smooth DOS is obtained by broadening the peaks, which we do following 
Ref.~\onlinecite{CostiBulla&C} through the transformation 
\begin{equation}
   \delta(\omega - \omega_{nm}) \rightarrow \frac{e^{-b^2/4}}
   {b\, \omega_{nm} \sqrt{\pi}} \exp{\left[ -
   \frac{(\ln{\omega} - \ln{\omega_{nm}})^2}{b^2} \right]},
\end{equation}
where $\omega_{nm} = E_n-E_m$ and $b=0.55$ for $\Lambda=2$.

\begin{figure}
\centerline{
\includegraphics[width=7.5cm]{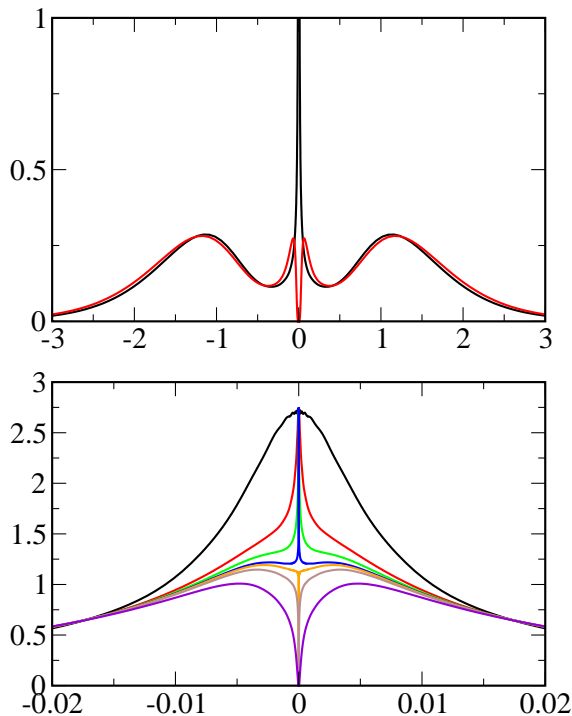} }
\caption{Impurity DOS in the presence of particle-hole symmetry across the fixed point. 
The temperature is set by the lenght of the chain; it is practically zero.  
In the upper panel we draw the DOS's well inside the Kondo screened phase $(J/J^*=0)$ and 
the unscreened one ($J/J^*=5.75$). Here $U=2$, $\Delta_0=U/(6\pi)$ and 
$J_*$ turns out to be $\simeq -0.0035$, all in units of half the conduction 
bandwidth. 
Notice the narrow peak which transforms into a narrow pseudo-gap.
In the lower panel we show in more detail the behavior of the low energy DOS across the UFP. 
(From top to bottom $J/J^*=0, 0.859, 0.945, 0.988, 1.002, 1.031, 1.146$)
\label{DOS-ph}
}
\end{figure}

In Fig.~\ref{DOS-ph} we show the outcome of the numerical calculation.
On the Kondo screened side of the UFP, the DOS shows a narrow Kondo resonance 
on top of a broader one. 
The height at the chemical potential is $\rho(0)=\rho_0$, as expected in a Kondo 
screened phase. 
On the contrary, in the unscreened side of the UFP, 
the narrow peak transforms into a narrow pseudo-gap 
within the broad resonance. Numerically we find that $\rho(\epsilon)\sim \epsilon^2$. 
As discussed before, 
this implies that the conventional behavior ${\cal I}m \Sigma(\epsilon) \sim \epsilon^2$ 
breaks down across the UFP. Exactly at the fixed point, both the narrow peak 
and the pseudo-gap disappear, leaving aside 
only the broad resonance. The calculated DOS at the chemical potential 
seems to be half of its non-interacting value, see Fig.~\ref{DOS-ph}. In other words our 
numerical results point to a DOS at the chemical potential which jumps across the UFP, 
being $\rho(0)=\rho_0$ everywhere in the Kondo-screened phase,  
$\rho(0)=0$ in the unscreened one, and $\rho(0)=\rho_0/2$ right at the UFP.

\section{Modeling the impurity density of states}
\label{Modelling}
\begin{figure}
\centerline{
\includegraphics[width=7.5cm]{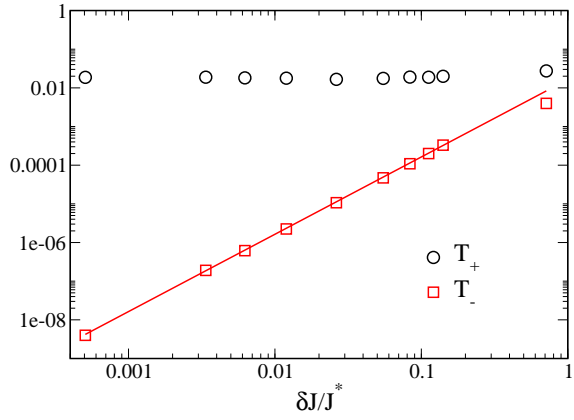} }
\caption{Fit values of $T_+$ and $T_-$ in the Kondo screened phase. The 
line is a quadratic fit, $T_-=A(\delta J)^2$. The Hamiltonian parameters have  
the same values as in Fig.~\ref{DOS-ph}. 
\label{T+-}
}
\end{figure}

It is possible to infer an analytical 
expression of the impurity DOS. First of all we notice that the values at the 
chemical potential in the screened and in the unscreened Kondo regimes are  
compatible with general scattering theory. In both phases the impurity 
has disappeared at low energy, either because it has been absorbed by the conduction sea 
or because $J$ has taken care of quenching the impurity spin and orbital degrees of freedom. 
This in turns means that what remains at low energy is just a potential scattering felt by  
the conduction electrons plus a local electron-electron interaction term.  
The on-shell $S$-matrix at the chemical potential has in general   
elastic and inelastic contributions 
(see Ref.~\onlinecite{Nozieres}). At zero temperature only the former survives. 
Since we considered just $s$-wave scattering, the elastic component of the $S$-matrix 
is given by:
\begin{equation}
S(0) = 1 - 2\pi i \rho_c T(0) = 1 - 2\,\pi\, \Delta_0 \,\rho(0),
\label{FL:elastic_S}
\end{equation}
where $\rho_c$ is the conduction electron DOS at the chemical potential per spin and 
band, and the $T$-matrix is defined through the conduction electron Green's function 
${\cal G}$ by  
\[
{\cal G} = {\cal G}_0 + {\cal G}_0\, T \, {\cal G}_0.
\]
On the other hand the $S$-matrix is related to the scattering phase shift by 
\begin{equation}
S(0) = {\rm e}^{2i\delta(0)}.
\label{FL:Svsdelta}
\end{equation}
In the Kondo screened phase, we know that $\delta(0)=\pi/2$ which, through 
(\ref{FL:Svsdelta}) and (\ref{FL:elastic_S}) implies $\rho(0) = 1/\pi\Delta_0$, namely 
its non-interacting value $\rho_0$. On the other hand, in the unscreened regime 
$\delta(0)=0$ hence $\rho(0)=0$, as we indeed find. It has been proposed that 
at the non-Fermi liquid fixed point of the overscreened $S=1/2$ two-channel Kondo model 
the $S$-matrix is instead purely inelastic\cite{A&L,L&M,Zarand}. 
That would imply a vanishing elastic contribution, $S(0)=0$ in (\ref{FL:elastic_S}), 
and in turns a DOS at the UFP
\begin{equation}
\rho(0)  = \frac{1}{2\pi\Delta_0} = \frac{1}{2} \rho_0,
\label{FL:DOS_FP}
\end{equation}
which is indeed compatible with our numerical results\cite{Zarand_grazie}.
Yet there is a difference between the UFP of our model, 
equivalently of the 2IKM, and the non-Fermi liquid fixed point of 
the $S=1/2$ two-channel Kondo model. While in the latter the specific heat has a singular 
temperature behavior right at the UFP, in our model it has a conventional linear 
behavior. The above observation suggests the following simple analytical expression of the 
low-energy impurity DOS:
\begin{equation}
\rho_{\pm}(\epsilon) = \frac{\rho_0}{2} 
\left( \frac{T_+^2}{\epsilon^2 + T_+^2} 
\pm \frac{T_-^2}{\epsilon^2 + T_-^2}\right),
\label{FL:modelDOS}
\end{equation}
where the plus sign refers to the Kondo screened phase and the minus to the 
unscreened one. The two energy scales have the same meaning as in the previous Section. 
In particular $T_-$ controls the deviations from the UFP, so that right at the UFP,   
when $T_-=0$, the DOS is 
\begin{equation}
\rho_*(\epsilon) = \frac{\rho_0}{2} \;
\frac{T_+^2}{\epsilon^2 + T_+^2}.  
\label{FL:modelDOS*}
\end{equation}
The model-DOS (\ref{FL:modelDOS}) also implies a model impurity Green's function 
in Matsubara frequencies:
\begin{equation}
G_\pm (i\epsilon_n) = 
\frac{1}{2\Delta_0}\left( 
\frac{T_+}{i\epsilon_n + iT_+ \,{\rm sign}\epsilon_n}
\pm \frac{T_-}{i\epsilon_n + iT_- \,{\rm sign}\epsilon_n}
\right).
\label{FL:G-ansatz}
\end{equation}
The fixed point Green's function, $G_* (i\epsilon_n)$, is identified by $T_-=0$. 
The impurity self-energy can then be extracted by the relation
\[
\Sigma_\pm(i\epsilon_n) = i\epsilon_n + i\Delta_0 \;{\rm sign}\,\epsilon_n 
- G_\pm(i\epsilon_n)^{-1}.  
\]
In particular, at low frequency we find that
\begin{equation}
i\epsilon_n - \Sigma_+(i\epsilon_n) \simeq i\epsilon_n \, \frac{\Delta_0}{2} 
\left( \frac{1}{T_+} + \frac{1}{T_-}\right),
\label{FL:Sigma+}
\end{equation}
in the Kondo screened phase, hence a standard linear behavior. On the 
contrary, in the unscreened regime the self-energy is singular 
\begin{equation}
i\epsilon_n - \Sigma_-(i\epsilon_n) \simeq - \frac{1}{i\epsilon_n} \,
\frac{2\Delta_0 T_+ T_-}{T_+-T_-}.
\label{FL:Sigma-}
\end{equation}
Finally, at the fixed point the self-energy is finite at zero frequency, being given by 
\begin{equation}
i\epsilon_n - \Sigma_*(i\epsilon_n) = i\Delta_0 
\, \frac{T_+ + 2 \epsilon_n}{T_+}.
\label{FL:Sigma*}
\end{equation}
We have checked that the model-self-energy gives indeed a good representation 
of the actual numerical results. In Fig.~\ref{T+-} 
we draw the fit values of $T_+$ and $T_-$ within the Kondo screened phase.

We can further test the consistency of the approach 
by invoking the scattering theory which, by 
the Friedel's sum rule,  allows us to identify the scattering phase shifts through: 
\begin{equation}
\delta(\epsilon) =   {\rm Im} \ln G(\epsilon+i0^+).
\label{FL:phase-shift}
\end{equation}
By means of our ansatz for the impurity Green's function (\ref{FL:G-ansatz}) 
we readily find that the expression of the low-energy phase-shifts is    
\begin{equation}
\delta_+(\epsilon) \simeq \frac{\pi}{2} + 
\frac{\epsilon}{2}\left(\frac{1}{T_+} + 
\frac{1}{T_-}\right)\equiv \frac{\pi}{2} + \alpha_+\epsilon \,
\label{FL:phase-shift KS}
\end{equation}
within the Kondo screened regime, and 
\begin{equation}
\delta_-(\epsilon) \simeq  
\epsilon \, \left(\frac{1}{T_+} + \frac{1}{T_-}\right)\equiv \alpha_-\epsilon,
\label{FL:phase-shift NKS}
\end{equation}
in the pseudo-gap unscreened phase, consistent with our starting assumption. 
Moreover, by the energy dependence of the phase shifts, we can calculate the impurity 
correction to the specific heat 
\begin{equation}
\frac{\delta C_V}{C_V} = \frac{\alpha_\pm}{\pi\rho_c}.
\label{FL:specific_heat}
\end{equation}

\section{Particle-hole symmetry breaking terms}
\label{Particle-hole symmetry breaking terms}
In this Section we analyse more in detail various symmetry breaking terms in the 
particle-hole channel. In particular we are going to consider the three following 
perturbations to the original Hamiltonian (\ref{Michele-ham}) with $\nu=0$: 
\bea
\delta H_{p-h} &=& \nu\, U\, n_d \equiv \frac{h_{p-h}}{2}\, n_d,
\label{p-hole}\\
\delta H_z &=& h_z\, T^z,
\label{p-h-z}\\
\delta H_x &=& h_x\, T^x.
\label{p-h-x}
\eea
The term (\ref{p-hole}) breaks particle-hole symmetry trying to occupy the impurity 
with $2-\nu$ electrons instead of two, see (\ref{Michele-ham}). The other terms, (\ref{p-h-z})  
and (\ref{p-h-x}), split the orbital degeneracy. 
Among them, only $\delta H_x$ is predicted to be relevant  
and wash out the fixed point, at least according to bosonization\cite{nostro}. 
Actually this looks a bit strange result if one invokes naively the argument of 
Ref.~\onlinecite{AL&J} to demonstrate the existence of an UFP in the absence 
of any particle-hole symmetry breaking term. This argument is based on the observation 
that, when O(2)$_{orbit}$ symmetry holds, the phase shifts in both orbital channels   
have to be equal, $\delta_1=\delta_2$. By general particle-hole symmetry, this 
further implies that $2\delta_1 = 2\delta_2= 0~mod(\pi)$. 
Since for $J\gg T_K>0$ we know that  $\delta_1=\delta_2=\pi/2$, while 
for $J\ll -T_K<0$, $\delta_1=\delta_2=0$, there must necessarily be a 
fixed point in between. 

Let us assume now that the $T^z$-term (\ref{p-h-z}) is 
present and follow Ref.~\onlinecite{AL&J} to demonstrate that the 
necessary condition for the existence of an intermediate 
fixed point does not hold anymore. Since (\ref{p-hole}) is absent, 
there is still a residual particle-hole symmetry according to which
\[
\delta_1+\delta_2=0~mod(\pi).
\]
If $\delta_1=-\delta_2$ then the two limiting cases, 
$\delta_1=\delta_2=0$ and $\delta_1=-\delta_2=\pi/2$, can be smoothly connected 
without requiring any critical point in between. This argument thus proves that 
an intermediate fixed point does not need to exist, yet it does not demonstrate its 
non-existence. Indeed we know by bosonization and we now show by NRG that both 
(\ref{p-h-z}) as well as (\ref{p-hole}) do not wash out the UFP. 
On the contrary a $T^x$-term (\ref{p-h-x}) does  
destabilize the fixed point, as shown later.

\begin{figure}
\centerline{
\includegraphics[width=7.5cm]{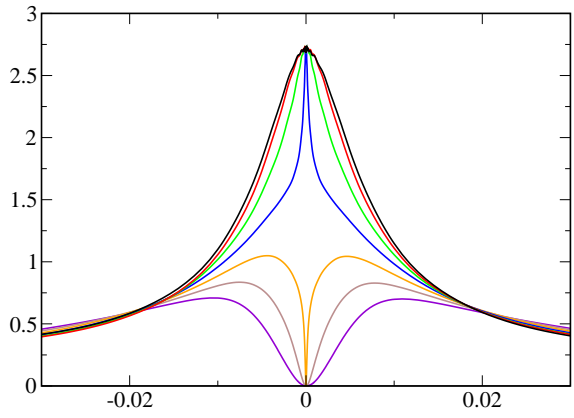} }
\caption{Impurity DOS across the UFP in the presence of a finite 
$\nu=0.05$ which breaks particle-hole 
symmetry. From top to bottom $J/J^*=0, 0.28, 0.57, 0.86, 1.14, 1.43, 1.71$. 
Notice that the DOS at the chemical potential is always finite, although very small 
hence not visible in the figure. $U$ and $\Delta_0$ have the same values as in 
Fig.~\ref{DOS-ph}.
\label{DOS-no-p-h}
}
\end{figure}

Let us go back to Eq.~(\ref{FL:elastic_S}) and try to guess how would it 
change in the presence of (\ref{p-hole}) and/or (\ref{p-h-z}). We have now 
to introduce one $S$-matrix for each channel, $S_a$ with $a=1,2$, satisfying 
\begin{equation}
{\cal R}e\, S_a(0) = \cos\,2\delta_a(0) = 1 - 2\,\pi\, \Delta_0 \,\rho_{a}(0). 
\label{FL:elastic_S_ph}
\end{equation}
Let us assume that, across the UFP, the zero-frequency phase shifts still 
jump by $\pi/2$. In other words, if we denote as   
\be
\delta_{-,a}(0) \equiv \delta_a,
\label{PH:def_delta}
\ee
the phase shift in the unscreened phase, in the Kondo-screened one the 
phase shift should be 
\[
\delta_{+,a}(0) = \delta_a + \frac{\pi}{2}.
\]
Through (\ref{FL:elastic_S_ph}) this would imply a jump of the 
DOS at the chemical potential given by 
\be
\rho_{+,a}(0)-\rho_{-,a}(0) = \frac{1}{\pi\Delta_0}\, \cos 2\delta_a = 
\rho_0\,\cos 2\delta_a.
\label{DOS-jump}
\ee 
The above scenario predicts that although the pseudo-gap in the unscreened phase is 
partly filled away from particle-hole symmetry, yet the DOS has a finite jump 
across the UFP. This is indeed confirmed by NRG. In Fig.~\ref{DOS-no-p-h} we
plot the DOS at fixed $\nu=0.05$, see Eq.~(\ref{p-hole}), across the UFP, clearly 
showing the jump. 
We notice that if only (\ref{p-hole}) is present, then 
$\delta_1=\delta_2$ 
in Eq.~(\ref{PH:def_delta}). If (\ref{p-hole}) is absent but 
(\ref{p-h-z}) is present, then $\delta_1=-\delta_2$, yet the behavior across 
the UFP is similar, which is the reason why we just show the results with finite $\nu$. 
This behavior is also compatible with the NRG result that the charge and $T^z$ 
Wilson ratios vanish around the UFP. Actually they all suggest that 
the model can absorb a chemical potential shift, equal or different in the two channels 
1 and 2, on a high energy scale, at least of order $T_+$, without having to 
modify what takes place at lower energies of order $T_-$: 
a kind of Anderson's compensation principle for our conserved quantities.    
Following these observations, we argue that the DOS for orbital $a=1,2$ 
in the presence of any of the two perturbations, (\ref{p-hole}) and (\ref{p-h-z}), 
assumed to be weak, can be modeled as 
\be
\rho_{\pm,a}(\epsilon) = \frac{\rho_{a}}{2}
\left[ \frac{T_+^2 + \mu_{\pm,a}^2}{(\epsilon+\mu_{\pm,a})^2+T_+^2}
\pm \cos 2\delta_{a}\; \frac{T_-^2}{\epsilon^2+T_-^2}\right],
\label{model-DOS-no-ph}
\ee
where again the plus refers to the Kondo screened phase, the minus to the 
unscreened one, $\rho_{a}=\rho_{+,a}(0)$ is the value of the DOS at 
the chemical potential in the screened regime, while 
\[
\mu_{\pm,a} = \pm T_+\, \sin 2\delta_a.
\]
According to the model DOS (\ref{model-DOS-no-ph}), the narrow peak and pseudo-gap 
remain pinned at the chemical potential, $\epsilon=0$, while only the broad 
resonance moves away from particle-hole symmetry.   

\begin{figure}
\centerline{
\includegraphics[width=7.5cm]{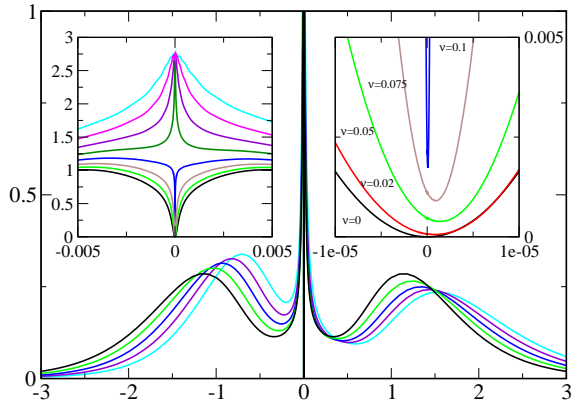} }
\caption{Impurity DOS upon increasing the 
strength of particle-hole symmetry breaking $\nu$ 
starting from the unscreened pseudo-gapped phase ($\nu=0, 0.05, 0.1$) up to the Kondo 
screened one ($\nu=0.15, 0.2$). 
In the left inset it is shown
the low energy part across the UFP (from top to bottom $\nu=0.2, 0.175, 0.15, 0.125, 0.1, 0.075, 0.05, 0$); 
notice the analogy with the p-h symmetric 
case in Fig.~\ref{DOS-ph}. In the right inset we explicitly show the gradual 
filling of the pseudo-gap upon increasing $\nu$. The values of $U$ and $\Delta_0$ 
are those of Fig.~\ref{DOS-ph}.
\label{DOS-vs-nu}
}
\end{figure}

Let us now study what happens if, starting from the particle-hole symmetric 
pseudo-gap phase we move away by increasing $\nu$, keeping all other Hamiltonian 
parameters fixed. As shown in Fig.~\ref{DOS-vs-nu}, 
$\nu$ is able to drive the model across the UFP. This result could be foreseen. 
Indeed $\nu$ forces the impurity to accomodate $2-\nu$ electrons. If $\nu=1$, 
the impurity tends to be singly-occupied. Therefore in the Kondo limit it behaves like 
a spin $S=1/2$ and pseudo-spin $T=1/2$ moment, which can be perfectly Kondo-screened and 
it is moreover stable with respect to little changes of $\nu$ with respect to $\nu=1$.
Hence, if the model is at $\nu=0$ in the pseudo-gap phase, it has to cross 
a fixed point to reach the large-$\nu$ Kondo screened regime.       
This behavior is quite interesting in connection with DMFT lattice calculations, 
since it implies that the lattice-model local critical regime, which reflects the 
single-impurity UFP,  
may also be attained by doping, as recently confirmed\cite{Capone-exE}.

\begin{figure}
\centerline{
\includegraphics[width=7.5cm]{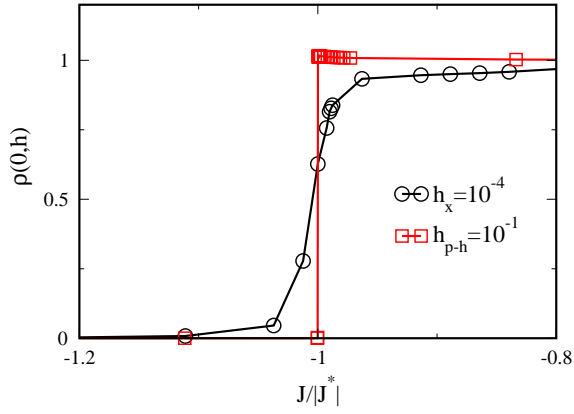} }
\caption{Comparison of the DOS values at the chemical potential as function of $J$ 
either in the presence of a finite particle-hole symmetry breaking 
$h_{p-h}$, $\rho(0,h_{p-h})$, or of a 
$T^x$ symmetry breaking $h_x$, $\rho(0,h_x)$, normalized to their values at $J=0$.  
Notice that $h_x$, although three order of magnitude smaller than 
$h_{p-h}$, washes out the DOS jump contrary to $h_{p-h}$. 
\label{DOS-vs-Tx}
}
\end{figure}

A completely different behavior occurs if we introduce instead a $T^x$ perturbation 
of the form (\ref{p-h-x}). Here, as expected, we do not find any jump of the DOS, 
as clear in Fig.~\ref{DOS-vs-Tx} where we 
compare the DOS at the chemical potential in the presence either of (\ref{p-hole}), 
$h_{p-h}\not = 0$, or (\ref{p-h-x}), $h_x\not =0$. This demonstrates that 
a perturbation in the particle-hole channel which breaks the orbital O(2) symmetry is 
relevant at the UFP, unlike (\ref{p-hole}) and (\ref{p-h-z}) which instead 
do not break the O(2)$_{orbit}$ symmetry.  

\medskip

\begin{figure}
\centerline{
\includegraphics[width=8cm]{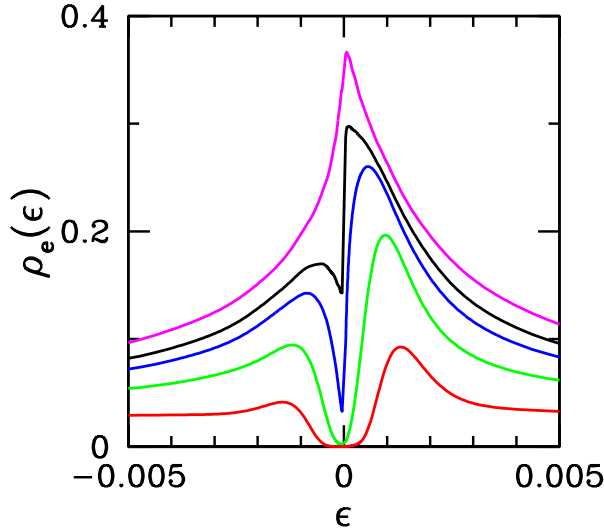} }
\caption{Impurity DOS of $d_{e\sigma}$, $\rho_e(\epsilon)$, 
for the AIM Eq.~(\ref{2HP-2IKM}). 
The different curves correspond from the top to the bottom to 
values of $\Delta_0=0.5, 0.47, 0.45, 0.4, 0.3$ with $t_\perp=0.05$ and $U=8$. 
These values correspond to $J_K=0.08,0.075,0.072,0.064,0.049$ and 
$J=4t_\perp^2/U=0.00125$. 
We notice the remnant of an asymmetric pseudo-gap of order $J$.  
\label{2Hubbard}
}
\end{figure}

\begin{figure}
\centerline{
\includegraphics[width=8cm]{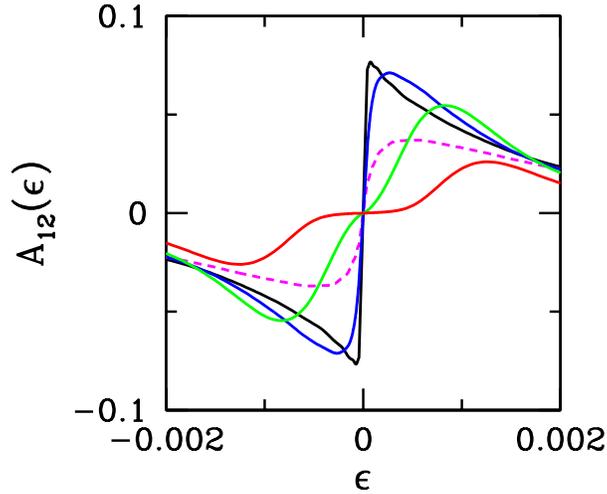} }
\caption{Off-diagonal spectral function, $A_{12}(\epsilon)$, with $t_\perp=0.05$. 
The different solid curves correspond from the top to the bottom for $\epsilon>0$ to 
values of $\Delta_0=0.47, 0.45, 0.4, 0.3$, while 
the dashed curve corresponds to $\Delta_0=0.5$. We notice that the low energy 
feature first moves towards zero energy when $\Delta_0$ increases from 
0.3 to 0.47, but from 0.47 to 0.5 it goes back again. Moreover, around 
$\Delta_0=0.47$, $A_{12}(\epsilon)$ is almost singular. 
\label{2Hubbard_bis}
}
\end{figure}

Finally let us discuss what happens in the AIM which corresponds within DMFT 
to two Hubbard planes coupled by a transverse hopping, Eq.~(\ref{H-Hubbard}) 
with $J=V=0$ plus the term (\ref{tperp}). We already noticed that 
$t_\perp$ plays an ambiguous role. It generates an antiferromagnetic exchange, 
$J=4t_\perp^2/U$, which may stabilize an UFP, but it also  
induces a relevant $T^x$ perturbation, see Eq.~(\ref{2HP-2IKM}).  
Since the UFP is never reachable, the model always flows to a Fermi liquid fixed point. 
In the presence of $t_\perp$ it is more appropriate to introduce 
the even and odd combinations of the orbitals 1 and 2:
\ba 
d_{e\sigma} &=& \frac{1}{\sqrt{2}}\left(d_{1\sigma} + d_{2\sigma}\right),\\
d_{o\sigma} &=& \frac{1}{\sqrt{2}}\left(d_{1\sigma} - d_{2\sigma}\right),
\ea
and correspondingly the even and odd conduction-electron scattering channels. 
According to what we said at the beginning of this Section, we expect  
the phase shifts $\delta_e=-\delta_o$ to be smooth functions of $J$.  
If there were no remnant of the UFP, the DOS's should 
simply show a resonance, the even channel above the chemical potential and the odd 
channel below it. In reality the behavior of the DOS remains strongly influenced 
by the UFP, even though never reachable. This is evident   
in Fig.~\ref{2Hubbard}, where we draw the DOS of $d_{e\sigma}$, $\rho_e(\epsilon)$, 
(the odd one is simply obtained by reflection around zero energy)  
at fixed $t_\perp$ upon varying the hybridization width $\Delta_0$. There is 
no point at which the DOS jumps at the chemical potential, yet a  
partly filled asymmetric pseudo-gap remains. 
In Fig.~\ref{2Hubbard_bis} we draw the low-energy difference between the even and odd DOS's, 
which is also the off-diagonal spectral function $A_{12}(\epsilon)$. 
$A_{12}(\epsilon)$ shows a low energy feature which has 
a non-monotonic behavior in $\Delta_0$ and 
almost develops into a singularity around $\Delta_0=0.47$.
We think that these results bring to the fore that $t_\perp$ alone is able to 
drive the model very close to the UFP.  
In other words the width of the critical region is larger than the energy scale 
which cut-off the fixed-point singularities, although both are generated by the 
same $t_\perp$.

\section{Conclusions}
\label{Conclusions}

In this work we have analysed the spectral properties 
of the two-orbital Anderson impurity model, Eq.~(\ref{Michele-ham}), 
which includes an exchange splitting $J$ which favors, if negative, 
a non-degenerate impurity configuration.  
This model was already shown in Ref.~\onlinecite{nostro} to 
posses a non-Fermi liquid fixed point which separates a phase where conventional 
Kondo screening takes place from an unscreened phase in which $J$ takes care of quenching 
the impurity degrees of freedom. 

The impurity density of states has the following behavior across the fixed point 
in the presence of particle-hole symmetry. 
In the Kondo screened phase it displays a conventional very narrow Kondo resonance 
on top of a broader resonance. On the contrary, in the unscreened phase 
a narrow pseudo-gap appears within the broad resonance. 
At the fixed point only the latter survives. Away from half-filling, the pseudo-gap 
remains pinned at the chemical potential, although it gets partly filled. Yet there 
is still a fixed point across which the density of states at the chemical potential jumps. 
Finally we have explicitly shown that the intermediate fixed point is unstable 
towards physical symmetry breaking fields, which include both particle-hole 
and particle-particle channels. The relevance of this impurity model for Dynamical Mean Field 
theory calculations has been already emphasized in Ref.~\onlinecite{nostro} and 
confirmed by Refs.~\onlinecite{Capone-exE,Han}. Here we would like to clarify some 
aspects in view of the newly discovered spectral properties.  

As discussed in Ref.~\onlinecite{nostro}, any lattice model which 
maps by DMFT onto the impurity model (\ref{Michele-ham}) plus 
(\ref{towards2IKM}) should encounter the unstable fixed point before the 
Mott transition, namely when the effective quasiparticle bandwidth becomes 
of the order of $|J|$. However the instability of the single-impurity fixed point 
should likely transform into a bulk instability through the DMFT self-consistency 
conditions. As we showed there are several competing physical instabilities 
around the fixed point, in the particle-hole and particle-particle channels. 
In the absence of nesting or van Hove singularities, 
we argued in Ref.~\onlinecite{nostro} that the 
particle-particle channel dominates, leading to a superconducting pocket just 
before the Mott transition, which has been indeed observed by DMFT\cite{Capone-exE}. 
However there might be physically relevant cases where 
those band-structure singularities occur, which would favor a uniform or modulated 
order-parameter in one of the particle-hole unstable channels. 
What should we expect upon moving away from these peculiar cases, for  
instance by doping? Clearly the band-structure singularities weakens   
upon doping. Yet the fixed point is not washed out away from particle-hole symmetry. 
We showed in fact that the pseudo-gap remains pinned at the chemical potential. 
We believe that this would result into a competition between particle-hole 
and particle-particle channels which gradually turns in favor of the latter, thus 
predicting a particle-hole order parameter which dies out  
upon doping in favor of a superconducting one.

Equally interesting is what we find for the Anderson impurity model which 
corresponds within DMFT to two Hubbard planes, with large in-plane coordination, 
coupled by an hopping term $t_\perp$, Eq.~(\ref{H-Hubbard}) 
with $J=V=0$ plus the term (\ref{tperp}).  
Here the physics is not as transparent as in the model (\ref{Michele-ham}),  
essentially because $t_\perp$ provides at the same time a mechanism for the existence 
of a fixed point as well as for its instability. However the numerical renormalization group 
results for the single-impurity show evidence that an almost critical region does exist, 
in spite of the fact that the non-trivial fixed point can never be attained.   
This suggests that the physics of the two coupled Hubbard planes close to the Mott transition 
may still be influenced by the 
single-impurity fixed point. 

Finally, we briefly comment what our results would imply for the model with 
conventional Hund's rules, see Eq.~(\ref{H_Hund}). This case in the 
absence of single-ion anisotropy corresponds to 
a Kondo screened regime where, as 
we showed in Fig.~\ref{FL:gamma}, the spin-triplet Cooper channel is attractive. 
By increasing the Hubbard $U$ in the lattice model, the Kondo temperature 
of the effective AIM decreases, which implies that the 
spin-triplet dimensionless scattering amplitude gradually also decreases, see 
Fig.~\ref{FL:gamma} for increasing $J>0$. This suggests that the instability 
towards spin-triplet superconductivity may actually be enhanced by strong correlations, 
compatible with recent DMFT calculations\cite{Han}. 
In addition we would expect that, in the presence of a single-ion 
anisotropy,  
$D>0$ in Eq.~(\ref{H_Hund}), the enhancement of 
spin-triplet superconductivity might be even more dramatic.

\begin{acknowledgments}
We are very grateful to G. Zar\'and,   
M. Sindel and C. Castellani for their useful advices.  
We thank M. Capone, P. Nozi\`eres, G.E. Santoro and E. Tosatti for 
helpful discussions. This work has been partly supported by MIUR COFIN and 
FIRB RBAU017S8R004.  
\end{acknowledgments}

\newpage
\appendix*

\section{Fermi liquid theory of the Anderson impurity model}
\label{FLFred}

In this Appendix we build up a Fermi liquid theory of our AIM closely following the 
conventional approach (see for instance Ref.~\onlinecite{Fred}). Our purpose is twofold. 
First the Fermi liquid theory provides a framework to analyse the 
NRG data. Moreover it allows to introduce within DMFT 
the concept of a local Fermi liquid description  
in addition to the conventional one, which refers instead to low frequency and momentum 
scattering amplitudes. 

Let us consider more generally a multi-orbital Anderson impurity model. 
As in our case, we assume that besides spin rotational symmetry 
also orbital degeneracy is preserved, so that the fully interacting 
impurity Green's functions are diagonal and independent either upon spin and orbital indices. 

The variation of the electron number with orbital symmetry $a$ and spin 
$\alpha$ associated with the presence of the impurity is given by\cite{Fred} 
\[
\Delta n_{a\alpha} = \oint \frac{ d z}{2\pi i} \, f(z) 
\frac{\partial}{\partial z} \ln G_{a\alpha}(z)
\]
where the integration contour encloses clockwise the real axis, $f(z)$ is the Fermi 
distribution function in the complex plane and $G_{a\alpha}$ 
the impurity single-particle Green's function. 
Since the Green's function has a 
branch cut on the real axis, the above expression is also equal to 
\begin{equation}
\Delta n_{a\alpha} = -\frac{1}{\pi} \int_{-\infty}^\infty d\epsilon \, 
\frac{\partial f(\epsilon)}{\partial \epsilon} 
\, {\rm Im} \ln G_{a\alpha}(\epsilon+i\delta) 
\label{FL:Delta-n}
\end{equation}
with $\delta$ an infinitesimal positive number. The impurity 
density of states is further determined through  
\begin{equation}
\rho_{a\alpha}(\epsilon) 
= -\frac{1}{\pi} {\rm Im}\, G_{a\alpha}\left(i\omega_n \to \epsilon + i\delta\right).
\label{FL:rho_d}
\end{equation}
If we introduce a source field in the Hamiltonian by 
\[
\delta \hat{H} = - \sum_{a\alpha} h_{a\alpha}\, n_{a\alpha},
\]
where 
\[
n_{a\alpha} = \sum_{\vect{k}} c^\dagger_{\vect{k}\,a\alpha}c^\dagga_{\vect{k}\,a\alpha} 
\,+\, d^\dagger_{a\alpha}d^\dagga_{a\alpha},
\]
then 
\[
G_{a\alpha}(i\omega_n)^{-1}\rightarrow 
i\omega_n + h_{a\alpha} - \Delta_{a\alpha}(i\omega_n,h_{a\alpha}) -
\Sigma_{a\alpha}(i\omega_n,\{h_{b\beta}\}),
\]
where 
\[
\Delta_{a\alpha}(i\omega_n,h_{a\alpha}) = 
\sum_{\vect{k}} |V_{\vect{k}}|^2 \, \frac{1}{i\omega_n - \epsilon_{\vect{k}} + h_{a\alpha}},
\]
is the hybridization function in the presence of the source. 
Therefore the derivative with respect to the external field of the 
variation of the electron number associated with the impurity is given by  
\begin{eqnarray}
\left(\frac{\partial \Delta n_{a\alpha}}{\partial h_{b\beta}}\right)_{h=0} &=& 
\int_{-\infty}^\infty \frac{d\epsilon}{\pi} \, 
\frac{\partial f(\epsilon)}{\partial \epsilon} \, {\rm Im}\, 
\left\{ G(\epsilon+i\delta)\phantom{\frac{G}{G}}\right.\nonumber \\
&& \left. \left[ \delta_{ab}\delta_{\alpha\beta} \left(1- 
\left(\frac{\partial \Delta(z)}{\partial z}\right)_{z=\epsilon+i\delta}\right)
- \left(\frac{\partial \Sigma_{a\alpha}(\epsilon+i\delta)}
{\partial h_{b\beta}}\right)_{h=0}\,\right]\right\},
\label{FL:var-Delta-n}
\end{eqnarray}
where $\Sigma_{a\alpha}(i\omega_n)$ is the impurity self-energy and 
we made use of 
\[
\left(\frac{\partial \Delta_{a\alpha}(z,h_{a\alpha})}{\partial h_{b\beta}}
\right)_{h=0} 
= \delta_{ab}\delta_{\alpha\beta}\, \frac{\partial \Delta(z)}{\partial z},
\]
being $\Delta(z)$ the hybridization function in the absence of $h$. 
On the other hand
\begin{equation}
\left(\frac{\partial \Sigma_{a\alpha}(i\omega_n)}{\partial h_{b\beta}}\right)_{h=0}   
= -\frac{1}{\beta}\sum_m \sum_{b\beta}\,\Gamma_{a\alpha,b\beta;b\beta,a\alpha}
(i\omega_n,i\epsilon_m;i\epsilon_m,i\omega_n) G(i\epsilon_m)^2
\, \left(1- 
\frac{\partial \Delta(i\epsilon_m)}{\partial i\epsilon_m}\right),
\label{FL:var-Sigma}
\end{equation}
where we used the property that, at $h=0$, the Green's function does not  
depend on $a$ and $\alpha$. The interaction vertex is the reducible one.

Let us assume that there exists a set of conserved operators
\ba
{\mathcal{M}}^{(i)} &=& 
\sum_{\vect{k}} \sum_{ab\alpha\beta} 
c^\dagger_{\vect{k}\,a\alpha}\, (\hat{M}^{(i)})_{ab}^{\alpha\beta}\, 
c^\dagga_{\vect{k}\,b\beta}\\ 
&&+ \sum_{ab\alpha\beta} d^\dagger_{a\alpha}
\, (\hat{M}^{(i)})_{ab}^{\alpha\beta}\, 
d^\dagga_{b\beta},
\ea
where $\hat{M}^{(i)}$ are hermitean  
matrices and the suffix $i$ 
identifies the particular conserved operator. For convenience we adopt the normalization 
${\rm Tr}\left(\hat{M}^{(i)}\cdot \hat{M}^{(i)}\right) = 1$. Then, if we add a source field
\[
\delta \hat{H} = -h^{(i)}\, {\mathcal{M}}^{(i)},
\]
we can use the basis which diagonalizes $\hat{M}^{(i)}$ and apply the above 
results to find the variation of $\langle {\mathcal{M}}^{(i)}\rangle$ 
associated with the presence of the impurity at first order 
in the applied field. Going back to the original basis, we would find the 
following expression of the difference $\delta \chi^{(i)}$ 
between the susceptibilities in the presence and absence of the impurity: 
\begin{eqnarray}
\delta \chi^{(i)} &=&  
\delta\, 
\left(\frac{\partial \langle {\mathcal{M}}^{(i)} \rangle }{\partial h^{(i)}}\right)_{h=0} 
\nonumber \\
&=& \int_{-\infty}^\infty \frac{d\epsilon}{\pi} \, 
\frac{\partial f(\epsilon)}{\partial \epsilon} \, 
{\rm Im}\, \left\{ G(\epsilon+i\delta)\phantom{\frac{G}{G}}\right. \nonumber \\
&& \left[ 1 - 
\left(\frac{\partial \Delta(i\epsilon)}{\partial i\epsilon}
\right)_{i\epsilon=\epsilon+i\delta}
+ \frac{1}{\beta}\sum_n \sum_{abcd}\sum_{\alpha\beta\gamma\delta} 
\Gamma_{b\beta,d\delta;c\gamma,a\alpha}(\epsilon+i\delta,i\epsilon_n;i\epsilon_n,\epsilon 
+ i\delta) \right. \nonumber \\
&& \left.\left. \left(\hat{M}^{(i)}\right)_{ab}^{\alpha\beta}\, 
\left(\hat{M}^{(i)}\right)_{cd}^{\gamma\delta}\, G(i\epsilon_n)^2\,
\left(1- 
\frac{\partial \Delta(i\epsilon_m)}{\partial i\epsilon_m}\right)\,
\right]\right\}.\nonumber\\
&&~~~~~\label{FL:chi}
\end{eqnarray}

Hereafter we drop the suffix $i$. One can demonstrate that the following 
Ward identities hold for the impurity 
\begin{eqnarray}
&&\left[ \Sigma(i\epsilon+i\omega) - 
\Sigma(i\epsilon)\right]\, M_{ab}^{\alpha\beta} \nonumber \\
&& = 
- \frac{1}{\beta}\sum_n \sum_{cd;\gamma\delta} 
\Gamma_{a\alpha,d\delta;c\gamma,b\beta}(i\epsilon+i\omega,i\epsilon_n;
i\epsilon_n + i\omega,i\epsilon)\nonumber\\
&&~~~~~~ M_{cd}^{\gamma\delta}\, G(i\epsilon_n+i\omega)\, 
G(i\epsilon_n) \, \left[i\omega - \Delta(i\epsilon_n+i\omega) 
+ \Delta(i\epsilon_n)\right].
\label{FL:Ward-identity}
\end{eqnarray}
It then follows that 
\begin{eqnarray}
\frac{\partial \Sigma(i\epsilon)}{\partial i\epsilon}\, M_{ab}^{\alpha\beta} &=& 
- \frac{1}{\beta}\sum_n \sum_{cd}\sum_{\gamma\delta} 
\Gamma_{a\alpha,d\delta;c\gamma,b\beta}(i\epsilon,i\epsilon_n;
i\epsilon_n,i\epsilon)\, M_{cd}^{\gamma\delta}\,G(i\epsilon_n)^2 \nonumber\\
&& 
- \lim_{i\omega\to 0} \frac{1}{\beta}\sum_n \sum_{cd;\gamma\delta} 
\Gamma_{a\alpha,d\delta;c\gamma,b\beta}(i\epsilon+i\omega,i\epsilon_n;
i\epsilon_n + i\omega,i\epsilon)\nonumber\\
&&~~~~~~ M_{cd}^{\gamma\delta}\, G(i\epsilon_n+i\omega)\, 
G(i\epsilon_n) \, \frac{ \left[- \Delta(i\epsilon_n+i\omega) 
+ \Delta(i\epsilon_n)\right]}{i\omega}\nonumber\\
&=& - \frac{1}{\beta}\sum_n \sum_{cd}\sum_{\gamma\delta} 
\Gamma_{a\alpha,d\delta;c\gamma,b\beta}(i\epsilon,i\epsilon_n;
i\epsilon_n,i\epsilon)\, M_{cd}^{\gamma\delta}\,G(i\epsilon_n)^2 
\left(1 - \frac{\partial \Delta(i\epsilon_n)}{\partial i\epsilon_n}\right)
\nonumber\\
&& +\int_{-\infty}^{\infty}\frac{d\epsilon'}{2\pi}\, 
\frac{\partial f(\epsilon')}{\partial\epsilon'}\, 
\sum_{cd;\gamma\delta} 
\Gamma_{a\alpha,d\delta;c\gamma,b\beta}(i\epsilon,
\epsilon'-i\delta';
\epsilon'+ i\delta',i\epsilon)\nonumber\\
&&
\left(M^{(i)}\right)_{cd}^{\gamma\delta}\,
\left(M^{(i)}\right)_{ba}^{\beta\alpha}\,
 G(\epsilon'+i\delta')\, 
G(\epsilon'-i\delta') \, {\rm Im}\left[
\Delta(\epsilon'-i\delta') - \Delta(\epsilon'+i\delta')\right].\nonumber \\
&& ~~~~\label{FL:Ward-identity-var}
\end{eqnarray}
Let us define the quantity
\begin{equation}
\bar{\rho}_* =  \int_{-\infty}^\infty \frac{d\epsilon}{\pi} \, 
\frac{\partial f(\epsilon)}{\partial \epsilon} 
\, {\rm Im}\left\{ G(\epsilon+i\delta)
\left[1 - \left(\frac{\partial \Delta(i\epsilon)}
{\partial i\epsilon}\right)_{i\epsilon\to \epsilon+i\delta} -
\left(\frac{\partial \Sigma(i\epsilon)}
{\partial i\epsilon}\right)_{i\epsilon\to \epsilon+i\delta} \right]
\right\},
\label{FL:def:z}
\end{equation}
which plays the role of the quasiparticle DOS at the chemical 
potential.
Then, through 
(\ref{FL:chi}), (\ref{FL:Ward-identity-var}) and (\ref{FL:def:z}), 
the following equation is readily found
\begin{eqnarray}
\bar{\rho}_* &=& 
\sum_{ab}\sum_{\alpha\beta} \bar{\rho}_* 
\, \left(\hat{M}^{(i)}\right)_{ba}^{\beta\alpha}
\, \left(\hat{M}^{(i)}\right)_{ab}^{\alpha\beta} \nonumber \\
&=& \delta \chi^{(i)} - \frac{1}{2\pi^2}
\int_{-\infty}^\infty d\epsilon \, d\epsilon'\,  
\frac{\partial f(\epsilon)}{\partial \epsilon} \,
\frac{\partial f(\epsilon')}{\partial \epsilon'} 
\, {\rm Im}\left\{G(\epsilon+i\delta)\left[\phantom{\frac{G}{G}}
\right.\right.
\nonumber \\
&&~~\sum_{cd;\gamma\delta} 
\Gamma_{a\alpha,d\delta;c\gamma,b\beta}(\epsilon+i\delta,
\epsilon'-i\delta';
\epsilon'+ i\delta',\epsilon+i\delta)\nonumber\\
&&\left.\left.
\left(M^{(i)}\right)_{cd}^{\gamma\delta}\,
\left(M^{(i)}\right)_{ba}^{\beta\alpha}\,
 G(\epsilon'+i\delta')\, 
G(\epsilon'-i\delta') \, {\rm Im}\left[
\Delta(\epsilon'-i\delta') - \Delta(\epsilon'+i\delta')\right]
\right]\right\}\nonumber\\
&=& \delta \chi^{(i)} + \frac{1}{2\pi}
\int_{-\infty}^\infty d\epsilon \, d\epsilon'\,  
\frac{\partial f(\epsilon)}{\partial \epsilon} \,
\frac{\partial f(\epsilon')}{\partial \epsilon'} 
\, \rho(\epsilon) \nonumber \\
&&~~\sum_{cd;\gamma\delta} 
\Gamma_{a\alpha,d\delta;c\gamma,b\beta}(\epsilon+i\delta,
\epsilon'-i\delta';
\epsilon'+ i\delta',\epsilon+i\delta)\nonumber\\
&&
\left(M^{(i)}\right)_{cd}^{\gamma\delta}\,
\left(M^{(i)}\right)_{ba}^{\beta\alpha}\,
 G(\epsilon'+i\delta')\, 
G(\epsilon'-i\delta') \, {\rm Im}\left[
\Delta(\epsilon'-i\delta') - \Delta(\epsilon'+i\delta')\right]\nonumber\\
&& ~~~~~~\label{FL:Ward-to-chi}
\end{eqnarray}
The last expression is obtained by noticing that only the imaginary 
part of $G(\epsilon+i\delta)$ contributes, where  
Im~$G(\epsilon+i\delta)= -\pi\rho(\epsilon)$. 
Eq.~(\ref{FL:Ward-to-chi}) allows to express any susceptibility 
to fields coupled to conserved quantities. 
If the hybridization function is smooth at low energies, then 
\[
\Delta(\epsilon'-i\delta') - \Delta(\epsilon'+i\delta') \simeq 2i\Delta_0,
\]
hence we can rewrite (\ref{FL:Ward-to-chi}) as follows
\begin{eqnarray}
\delta \chi^{(i)} &=& \bar{\rho}_* 
\, \left[ 1 - 
\frac{\Delta_0}{\bar{\rho}_*\pi}
\int_{-\infty}^\infty d\epsilon \, d\epsilon'\,  
\frac{\partial f(\epsilon)}{\partial \epsilon} \,
\rho(\epsilon)\, \frac{\partial f(\epsilon')}{\partial \epsilon'} \right.
\nonumber \\
&&~~\sum_{cd;\gamma\delta} 
\Gamma_{a\alpha,d\delta;c\gamma,b\beta}(\epsilon+i\delta,
\epsilon'-i\delta';
\epsilon'+ i\delta',\epsilon+i\delta)\nonumber\\
&&
\left. \left(M^{(i)}\right)_{cd}^{\gamma\delta}\,
\left(M^{(i)}\right)_{ba}^{\beta\alpha}\,
 G(\epsilon'+i\delta')\, 
G(\epsilon'-i\delta')\right], \nonumber \\
&\equiv& \bar{\rho}_* \, \left[ 1 - A_i\right], \label{FL:chi-final}
\end{eqnarray}
which allows to identify local Landau $A$-parameters through
\begin{eqnarray}
A_i &=& 
\frac{\Delta_0}{\bar{\rho}_*\pi}
\int_{-\infty}^\infty d\epsilon \, d\epsilon'\,  
\frac{\partial f(\epsilon)}{\partial \epsilon} \,
\rho(\epsilon)\, \frac{\partial f(\epsilon')}{\partial \epsilon'} 
\nonumber \\
&&~~\sum_{cd;\gamma\delta} 
\Gamma_{a\alpha,d\delta;c\gamma,b\beta}(\epsilon+i\delta,
\epsilon'-i\delta';
\epsilon'+ i\delta',\epsilon+i\delta)\nonumber\\
&&
\left. \left(M^{(i)}\right)_{cd}^{\gamma\delta}\,
\left(M^{(i)}\right)_{ba}^{\beta\alpha}\,
 G(\epsilon'+i\delta')\, 
G(\epsilon'-i\delta')\right]. 
\label{FL:def:A}
\end{eqnarray}
The above expression is quite general but simplifies substantially when the 
imaginary part of the impurity self-energy vanishes at low real frequency. In this case 
\[
G(i\epsilon_n\to \pm i 0^+) = \frac{1}{-\epsilon_d  
\pm i \, \Delta_0},
\]
where $\epsilon_d = \epsilon^{(0)}_d + {\rm Re}\Sigma(0)$ 
is the actual position of the $d$-resonance. Then, through (\ref{FL:rho_d}),  
\begin{equation}
\rho(0) = \frac{1}{\pi} \frac{\Delta_0}{\epsilon_d^2 + \Delta_0^2} 
= \frac{\Delta_0}{\pi} G(i0^+)G(i0^-).
\label{FL:rho_d(0)}
\end{equation} 
Analogously 
\[
\bar{\rho}_* = \frac{\rho(0)}{Z},\;\; \frac{1}{Z} = 1 
 - \left(\frac{\partial \Sigma(i\epsilon)}
{\partial i\epsilon}\right)_{i\epsilon\to i0^+}
\]
hence 
\begin{equation}
A_i = \sum_{abcd}\sum_{\alpha\beta\gamma\delta} 
\left[ Z^2 \, \bar{\rho}_*\, \Gamma_{a\alpha,d\delta;c\gamma,b\beta}(0,0;0,0)\right] \,
\left(\hat{M}^{(i)}\right)_{ba}^{\beta\alpha}\,
\left(\hat{M}^{(i)}\right)_{cd}^{\gamma\delta},
\label{FL:def:AFL}
\end{equation}
which is the more conventional expression of the Landau parameters\cite{Fred}. Although the 
above equation is a particular case of the general one (\ref{FL:def:A}), 
to simplify the notations in what follows we will use (\ref{FL:def:AFL}) as 
a short-hand expression of (\ref{FL:def:A}). 

\subsection{Application to the twofold orbitally degenerate AIM} 
Let us now apply the above results to our model. An incoming pair can be 
a spin-triplet orbital-singlet, with a scattering vertex at zero incoming and 
outgoing frequencies given by 
\[
\Gamma^1 \rightarrow   \Gamma_{1\sigma,2\sigma;2\sigma,1\sigma},\;\;  
 \frac{1}{2}\Gamma_{1\sigma,2\, -\sigma;2\, - \sigma,1\sigma} 
-\frac{1}{2}\Gamma_{1\sigma,2\, -\sigma;1\, - \sigma,2\sigma}. 
\]
Here 1 and 2 label the two orbitals with $T^z=+1/2$ and $T^z=-1/2$, respectively. 
Alternatively it can be a spin-singlet orbital-triplet with $T^z=0$, with scattering 
vertex 
\[
\Gamma^0_0 \rightarrow     
\frac{1}{2}\Gamma_{1\sigma,2\, -\sigma;2\, - \sigma,1\sigma} +  
\frac{1}{2}\Gamma_{1\sigma,2\, -\sigma;1\, - \sigma,2\sigma},
\]
or with $T^z=\pm 1$, in which case 
\[
\Gamma^0_{\pm} \rightarrow 
\Gamma_{1\sigma,1\, -\sigma;1\, - \sigma,1\sigma},\;\;\; 
\Gamma_{2\sigma,2\, -\sigma;2\, - \sigma,2\sigma}.
\]
In reality it is more convenient to introduce the dimensionless scattering vertices:
\be
\begin{array}{lcl}
\A^1 &=& Z^2 \, \bar{\rho}_*\, \Gamma_1 = Z\, \rho(0)\, \Gamma^1,\\
\A^0_0 &=& Z^2 \, \bar{\rho}_*\, \Gamma^0_0 = Z\, \rho(0)\, \Gamma^0_0,\\
\A^0_{\pm} &=& Z^2 \, \bar{\rho}_*\, \Gamma^0_{\pm} = Z\, \rho(0)\, \Gamma^0_{\pm}.\\
\end{array}
\label{FL:dimen-A}
\ee
As we previously showed, only the susceptibilities of conserved quantities can be 
expressed in terms of the Landau parameters (\ref{FL:def:AFL}), which are 
simply connected with the scattering vertices at zero frequency. Yet we can 
still define Landau parameters for non-conserved quantities, which, although 
do not serve to calculate susceptibilities, may provide a qualitative estimate of 
their magnitude. Therefore we are going to introduce the Landau parameters  
for the charge, $A_C$, spin $A_S$, the $z$-component of the pseudo-spin $\vec{T}$, 
$A^{||}_T$, all being related to conserved quantities, but also 
for the $x$ and $y$ components of $\vec{T}$, $A^{\perp}_T$, as well as for 
the spin-orbital components, $A^{||}_{ST}$ and $A^{\perp}_{ST}$. In terms of the 
dimensionless amplitudes (\ref{FL:dimen-A}) they 
can be shown, after some lengthy algebra, to have the following expressions: 
\begin{eqnarray}
A_C &=& \frac{1}{4}\left(6\A^1 + 2\A^0_0 + 4\A^0_{\pm}\right),
\label{Ac}\\
A_S &=& \frac{1}{4}\left(2\A^1 - 2\A^0_0 - 4\A^0_{\pm}\right),
\label{AS}\\
A^{||}_T &=& \frac{1}{4}\left(-6\A^1 - 2\A^0_0 + 4\A^0_{\pm}\right),
\label{ATz}\\
A^{\perp}_T &=& \frac{1}{4}\left(-6\A^1 + 2\A^0_0 \right),
\label{ATxy}\\
A^{||}_{ST} &=& \frac{1}{4}\left(-2\A^1 + 2\A^0_0 - 4\A^0_{\pm}\right),
\label{FL:ASTz}\\
A^{\perp}_{ST} &=& \frac{1}{4}\left(-2\A^1 - 2\A^0_0 \right).
\label{FL:ASTxy}
\end{eqnarray}
Let us consider several possible cases.
\begin{itemize}
\item If $J=0$, SU(4) symmetry holds. Then $\A^1=\A^0_0=\A^0_{\pm}=A$, leading 
to 
\begin{eqnarray*}
A_C &=& 3\,A,\\
A_S &=& A^{||}_T = A^{\perp}_T = A^{||}_{ST}=
A^{\perp}_{ST} = -A.
\end{eqnarray*}
In the $s$-$d$ limit, when the AIM maps onto an SU(4) Kondo model, the 
charge compressibility in negligible, leading to $3A =1$. 
The Wilson ratios for the conserved quantities are defined trough
\[
R_i = \frac{\delta \chi^{(i)}}{\chi_0} \frac{C_V}{\delta C_V} = 1-A_i,
\]
where $\delta \chi^{(i)}$ has been defined in (\ref{FL:chi}), 
$\chi_0=\rho_c$ and $C_V$ are respectively the conduction-electron susceptibility 
and specific heat in the absence of the impurity, and 
\[
\delta C_V = \frac{\bar{\rho}_*}{\rho_c}\, C_V,
\]
is the variation of the specific heat due to the impurity. 
Hence all Wilson ratios have an universal value, 
\be
R_S=R_T=R_{ST}= 1-A = 4/3,
\label{FL:Wilsonratio}
\ee
in agreement with Conformal Field Theory. 
\item If $J\gg T_K>0$ the impurity gets 
frozen in the Kondo limit into a spin S=1. Then both $A_C=1$ and $A^{||}_T=1$, 
which implies 
\begin{eqnarray*}
\A^0_{\pm} &=& 1, \\
\A^0_0 &=& - 3\, \A^1.
\end{eqnarray*} 
However one expects that, being the spin-triplet an orbital singlet, the 
SU(2) orbital symmetry gets restored at the fixed point, much in the same way as 
spin anisotropy is irrelevant at the Kondo fixed point. This further implies that 
\[
\A^0_0 = - 3\A^1 = 1,
\]
namely $A_S = -5/3$, with a Wilson ratio $R_S = 8/3$, in agreement with known results.   
\item Let us now suppose to be close to the UFP within the Kondo screened regime. 
As usual the charge degrees of freedom are suppressed already below $U$,
so that we can still assume $A_C=1$.   
Moreover we expect that the spin and the orbital degrees of freedom related to $T^z$ 
get quenched below $T_+$, while the remaining ones only below $T_-\ll T_+$. Therefore 
at very low temperatures $T< T_-$, we can safely assume that 
\[
T_- \, \delta\chi_S \sim T_-\, \delta \chi^{||}_T\sim \frac{T_-}{T_+} \sim 0,
\]
namely $A_S=A^{||}_T=1$. As a result we find that 
\begin{eqnarray}
\A^0_{\pm} &=& \A^1 = 1,\label{FL:FP1}\\
\A^0_{0} &=& -3.\label{FL:FP2}
\end{eqnarray}
Eq.~(\ref{FL:FP2}) implies a strongly attractive $s$-wave singlet channel. The other 
Landau parameters are thus given by 
\begin{eqnarray}
A^{\perp}_T &=& A^{||}_{ST} = -3,\label{FL:relevantA}\\
A^{\perp}_{ST} &=& 1.\label{FL:irrelevantA}
\end{eqnarray}
This further proves that the fixed point is equally unstable 
in the $s$-wave Cooper channel $\Gamma^0_0$, as well as in the 
$T^x$, $T^y$ and $\vec{S}\,T^z$ particle-hole channels. 

We finally notice that, although the Landau $A$-parameters would suggest that 
the susceptibilities in the unstable channels, all of which 
correspond to non-conserved quantities, diverge as $1/T_-$, in reality they only diverge 
logarithmically\cite{Varma,Gan,AL&J}. 
This is not incompatible with Fermi liquid theory, which allows to express in terms of the 
$A$-parameters only those response functions related to conserved quantities.   
   
\end{itemize} 

\medskip

Let us now use our model self-energy to extract some additional information.
Through Eq.~(\ref{FL:Sigma+}), we find that in the Kondo screened regime  
the expression (\ref{FL:def:AFL}) holds with a quasiparticle residue  
\begin{equation}
\frac{1}{Z} = \frac{\Delta_0}{2}\left(\frac{1}{T_+} + \frac{1}{T_-}\right).
\label{FL:z-FL}
\end{equation}
Indeed $Z\sim 2T_-/\Delta_0 \to 0$ upon approaching the unstable fixed point. 

On the contrary, the general expression (\ref{FL:def:A}) has to be used inside the 
non-Kondo screened pseudo-gap phase. Through Eq.~(\ref{FL:G-ansatz}) for $G_-(i\epsilon_n)$ 
we find that at low frequency 
\[
G_-(\epsilon+i\delta)\, G_-(\epsilon-i\delta) \simeq \frac{1}{4\Delta_0^2} 
\frac{\epsilon^2 \left(T_+-T_-\right)^2}{T_+^2 T_-^2} 
\simeq 
\frac{\pi}{2\Delta_0} \, \rho_{-}(\epsilon)\, \frac{T_+-T_-}{T_+ + T_-}.
\]
By Eq.~(\ref{FL:Sigma-}) the quasiparticle DOS 
at the chemical potential turns out to be finite,  
\begin{equation}
\bar{\rho}_* = \frac{1}{\pi} \frac{T_+ + T_-}{T_+ T_-},
\label{FL:rho*}
\end{equation}
even though the impurity DOS vanishes. In conclusion, within the pseudo-gap phase 
the Landau parameters have the following expression
\begin{eqnarray}
A_i &=& 
\frac{\pi}{2} \frac{T_+ T_- \left(T_+-T_-\right)}{\left(T_++T_-\right)^2}\; 
\int_{-\infty}^\infty d\epsilon \, d\epsilon'\,  
\frac{\partial f(\epsilon)}{\partial \epsilon} \,
\rho(\epsilon)\, \frac{\partial f(\epsilon')}{\partial \epsilon'}\, 
\rho(\epsilon') 
\nonumber \\
&&~~\sum_{cd;\gamma\delta} 
\Gamma_{a\alpha,d\delta;c\gamma,b\beta}(\epsilon+i\delta,
\epsilon'-i\delta';
\epsilon'+ i\delta',\epsilon+i\delta)\nonumber\\
&&
\left. \left(M^{(i)}\right)_{cd}^{\gamma\delta}\,
\left(M^{(i)}\right)_{ba}^{\beta\alpha} \right]. 
\label{FL:def:ANFL}
\end{eqnarray}
In spite of the anomalous impurity Green's 
function, the low-energy behavior should still be described within a local 
Fermi liquid scenario by finite Landau parameters $A_i$'s. 
Therefore, since the impurity DOS vanishes quadratically in the pseudo-gap 
phase, then the scattering vertices must display a singular behavior  
\[
\Gamma(\epsilon,
\epsilon';
\epsilon',\epsilon) \sim \frac{1}{
(\epsilon+\epsilon')^4},
\]
to compensate for the vanishing DOS's and provide finite $A$'s.

\newpage
\begingroup
\squeezetable
\begin{table}
\caption{Energies $E$ of the low energy levels 
and their degeneracy $deg$ at the unstable fixed point. The levels are 
labeled by the quantum numbers $Q$, half of the deviation of the number of electrons 
with respect to the ground state, $S$, total spin, and $T^z$, total $z$-component of the 
pseudo-spin. The value $x$ is the prediction of Conformal Field Theory for the 
two-impurity Kondo model\cite{AL&J}. 
Notice the anomaly of the member within the $(1/2,1/2,1/2)$ multiplets 
identified by a $^*$, which was also found in Ref.~\onlinecite{AL&J}. 
There an explanation for the discrepancy was proposed.}
\label{table} 
   \begin{tabular}{|ccc|c|c|c|}
      \hline
      $Q$ & $T^z$ & $S$ & $x$ & $E$ & deg\\
      \hline
      0   & 0   & 0   & 0   & 0.00000 & 1\\
      \hline
      $\frac{1}{2}$ & $\frac{1}{2}$ & $\frac{1}{2}$ & $\frac{3}{8}$
       & 0.37260 & 8\\
      \hline
      0   & 0   & 1   & $\frac{1}{2}$ & 0.49615 & 3\\
      0   & 1   & 0   & $\frac{1}{2}$ & 0.49583 & 2\\
      1   & 0   & 0   & $\frac{1}{2}$ & 0.49631 & 2\\
      \hline
      $\frac{1}{2}$ & $\frac{1}{2}$ & $\frac{1}{2}$ & 
        $\frac{7}{8}$ & 0.88021 & 8\\
      \hline
      0   & 0   & 0   & 1   & 0.99714 & 1\\
          &     &     &     & 1.00216 & 1\\
          &     &     &     & 1.00311 & 1\\
      0   & 0   & 1   & 1   & 1.00279 & 3\\
      0   & 1   & 1   & 1   & 1.00248 & 6\\
      1   & 0   & 1   & 1   & 1.00295 & 6\\
      1   & 1   & 0   & 1   & 1.00264 & 4\\
      \hline
      $\frac{1}{2}$ & $\frac{1}{2}$ & $\frac{1}{2}$ & 1+$\frac{3}{8}$ 
        & 1.38880 & 8\\
          &     &     &     & 1.38945 & 8\\
          &     &     &     & 1.51556$^*$ & 8\\
      $\frac{1}{2}$ & $\frac{1}{2}$ & $\frac{3}{2}$ & 1+$\frac{3}{8}$ 
       & 1.38924 & 16\\
      $\frac{1}{2}$ & $\frac{3}{2}$ & $\frac{1}{2}$ & 1+$\frac{3}{8}$ 
       & 1.38859 & 8\\
      $\frac{3}{2}$ & $\frac{1}{2}$ & $\frac{1}{2}$ & 1+$\frac{3}{8}$ 
       & 1.38957 & 8\\
      \hline
      0   & 0   & 0   & 1+$\frac{1}{2}$ & 1.55944 & 1\\
      0   & 0   & 1   & 1+$\frac{1}{2}$ & 1.50195 & 3\\
          &     &     &               & 1.55863 & 3\\
          &     &     &               & 1.55983 & 3\\
          &     &     &               & 1.60582 & 3\\
      0   & 1   & 0   & 1+$\frac{1}{2}$ & 1.50141 & 2\\
          &     &     &               & 1.55943 & 2\\
          &     &     &               & 1.60467 & 2\\
      0   & 1   & 1   & 1+$\frac{1}{2}$ & 1.55904 & 6\\
      1   & 0   & 0   & 1+$\frac{1}{2}$ & 1.50222 & 2\\
          &     &     &               & 1.55883 & 2\\
          &     &     &               & 1.60636 & 2\\
      1   & 0   & 1   & 1+$\frac{1}{2}$ & 1.55964 & 6\\
      1   & 1   & 1   & 1+$\frac{1}{2}$ & 1.55923 & 12\\
      \hline
   \end{tabular}
\end{table}
\endgroup


\begin{thebibliography}{30}
\bibitem{nota} In Ref.~\onlinecite{nostro} there is a misprint concerning 
this point. Indeed the response to $ST^x$ and $ST^y$ fields 
are not singular, as erroneously written in that reference.
\end{thebibliography}
\end{document}